\documentclass[10pt,journal,compsoc]{IEEEtran}
    
\usepackage{amssymb}
\usepackage{graphicx}
\usepackage{epstopdf}
\usepackage{gensymb}
\usepackage{color}
\usepackage{amsmath}
\usepackage{enumerate}
\usepackage{cite}
\usepackage{comment}
\usepackage{amsthm}
\usepackage{enumitem}
\theoremstyle{definition}

\usepackage{float}
\usepackage{subfigure}
\usepackage{soul}

\usepackage[framemethod=TikZ]{mdframed}
\usepackage{capt-of}
\usepackage{algorithmicx}
\usepackage[ruled]{algorithm}
\usepackage{algpseudocode}
\alglanguage{pseudocode}

\usepackage{bbding}
\usepackage{wasysym}

\usepackage{array}
\usepackage{booktabs}
\usepackage{multirow}

\usepackage{threeparttable} 
\usepackage{hyperref}

\usepackage{mathtools}

\usepackage{url}
\urldef{\mailsa}\path|{alfred.hofmann, ursula.barth, ingrid.haas, frank.holzwarth,|
	\urldef{\mailsb}\path|anna.kramer, leonie.kunz, christine.reiss, nicole.sator,|
	\urldef{\mailsc}\path|erika.siebert-cole, peter.strasser, lncs}@springer.com|

\definecolor{myblue}{rgb}{0,0,0}

\begin{document}

\title{Quantization Backdoors to Deep Learning Commercial Frameworks}

\author{

Hua Ma\IEEEauthorrefmark{1}, Huming Qiu\IEEEauthorrefmark{1}, Yansong Gao\IEEEauthorrefmark{2} ({\it Member, IEEE}), Zhi Zhang, Alsharif Abuadbba, Minhui Xue, \\ Anmin Fu, Jiliang Zhang, Said F. Al-Sarawi ({\it Senior Member, IEEE}), Derek Abbott ({\it IEEE Fellow})

\IEEEcompsocitemizethanks{\IEEEcompsocthanksitem \IEEEauthorrefmark{1}H.~Ma and \IEEEauthorrefmark{1}Q.~Li contributed equally to the study.}

\IEEEcompsocitemizethanks{\IEEEcompsocthanksitem \IEEEauthorrefmark{2} Corresponding author.}

\IEEEcompsocitemizethanks{\IEEEcompsocthanksitem H.~Ma, S.~Al-Sarawi, and D.~Abbott are with the School of Electrical and Electronic Engineering, The University of Adelaide, Australia. H.~Ma is also with Data61, CSIRO.
\{hua.ma;said.alsarawi;derek.abbott\}@adelaide.edu.au}

\IEEEcompsocitemizethanks{\IEEEcompsocthanksitem H.~Qiu and A.~Fu are with the School of Computer Science and Engineering, Nanjing University of Science and Technology, China. \{120106222682;fuam\}@njust.edu.cn}

\IEEEcompsocitemizethanks{\IEEEcompsocthanksitem Y.~Gao is with Data61, CSIRO, Australia. This work was mainly done when he was with the School of Computer Science and Engineering, Nanjing University of Science and Technology, China. gao.yansong@hotmail.com}

\IEEEcompsocitemizethanks{\IEEEcompsocthanksitem Z.~Zhang is with the University of Western Australia, Perth, Australia. zhi.zhang@uwa.edu.au}

\IEEEcompsocitemizethanks{\IEEEcompsocthanksitem A.~Abuadbba and M.~Xue are with Data61, CSIRO, Australia. \{sharif.abuadbba;jason.xue\}@data61.csiro.au}

\IEEEcompsocitemizethanks{\IEEEcompsocthanksitem J.~Zhang is with the College of Semiconductors (College of Integrated Circuits) at Hunan University, China.
zhangjiliang@hnu.edu.cn}

}

\IEEEtitleabstractindextext{		
\begin{abstract}
Due to their low latency and high privacy preservation, there is currently a burgeoning demand for deploying deep learning (DL) models on ubiquitous edge Internet of Things (IoT) devices. However, DL models are often large in size and require large-scale computation, which prevents them from being placed directly onto IoT devices, where resources are constrained, and 32-bit floating-point (float-32) operations are unavailable. Commercial framework (i.e., a set of toolkits) empowered model quantization is a pragmatic solution that enables DL deployment on mobile devices and embedded systems by effortlessly post-quantizing a large high-precision model (e.g., float-32) into a small low-precision model (e.g., int-8) while retaining the model inference accuracy. However, their usability might be threatened by security vulnerabilities.

This work reveals that standard quantization toolkits can be abused to activate a backdoor. We demonstrate that a full-precision backdoored model which does not have any backdoor effect in the presence of a trigger---as the backdoor is dormant---can be activated by (i) TensorFlow-Lite (TFLite) quantization, the only \textit{product-ready} quantization framework to date, and (ii) the \textit{beta released} PyTorch Mobile framework. In our experiments, we employ three popular model architectures (VGG16, ResNet18, and ResNet50), and train each across three popular datasets: MNIST, CIFAR10 and GTSRB. We ascertain that all trained float-32 backdoored models exhibit no backdoor effect \textit{even in the presence of trigger inputs}. Particularly, four influential backdoor defenses are evaluated, and they fail to identify a backdoor in the float-32 models. When each of the float-32 models is converted into an int-8 format model through the standard TFLite or PyTorch Mobile framework's post-training quantization, the backdoor is activated in the quantized model, which shows a stable attack success rate close to 100\% upon inputs with the trigger, while it usually behaves upon non-trigger inputs. This work highlights that a stealthy security threat occurs when an end-user utilizes the on-device post-training model quantization frameworks, informing security researchers of a cross-platform overhaul of DL models post-quantization even if these models pass security-aware front-end backdoor inspections. Significantly, we have identified Gaussian noise injection into the malicious full-precision model as an easy-to-use preventative defense against the PQ backdoor. The attack source code is released at  \url{https://github.com/quantization-backdoor}.
\end{abstract}

}

\maketitle

\section{Introduction}\label{sec:Intro}
Deep learning (DL) empowers a wide range of applications, such as computer vision and natural language processing.
Traditionally, DL models are trained and hosted in the cloud for inference tasks. While the cloud provides massive computing power for model training, it does introduce latency for model inference (each inference requires network connectivity for a round-trip to the cloud), posing a severe challenge to real-time applications (e.g., autonomous driving).
In addition, users' personal data have to be submitted to the cloud for recommendation or inference tasks, putting data privacy at risk.  To address these key constraints, on-device DL (also called TinyML~\cite{warden2019tinyml}) empowers end-users to perform model inference directly in ubiquitous IoT devices.
Compared to in-cloud DL, on-device DL voids the requirement of connecting local data to the cloud and preserves user data privacy in this context.

\begin{figure}[t]
	\centering
	\includegraphics[trim=0 0 0 0,clip,width=0.40 \textwidth]{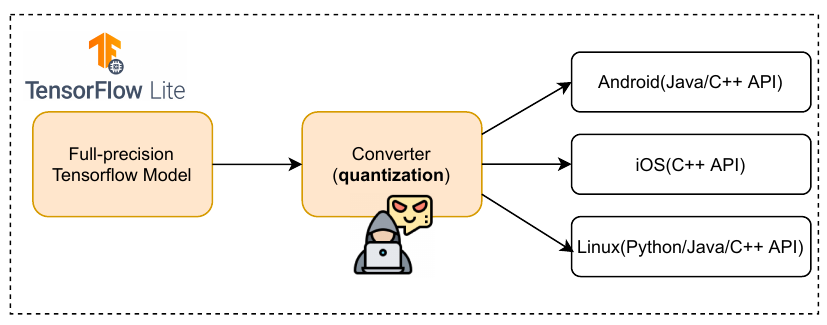}
	\caption{A general pipeline of the TFLite post-training quantization framework~\cite{karthikeyan2018machine}, where the converter performing the model quantization is exploited to wake up a dormant backdoor within the full-precision model. Dormant here means that the backdoor of the full-precision model \textit{cannot be activated even in the presence of trigger inputs}.}
	\label{fig:TFLite}
	\vspace{-2mm}
\end{figure}

Recently, on-device DL models have been deployed to various IoT devices such as microcontrollers (MCUs), the smallest computing platforms present almost everywhere.\footnote{The world is estimated to have over 250 billion MCUs in 2020 and the number is increasing rapidly~\cite{ICInsights}.} 
As IoT devices are resource-constrained, on-device DL cannot directly deploy a DL model trained from the cloud onto the devices. Particularly, a model is usually trained using a floating-point precision format (e.g., a full precision in float-32) and can be much larger (e.g., a few hundred MiB) than the memory capacity (within a few MiB) of most IoT devices. Further, the model requires computing-intensive floating-point operations (FLOPS), that are unlikely to be supported in low-power IoT devices (e.g., drones and smart watches).
To make DL ubiquitous, a practical solution called \textit{model quantization} has been proposed~\cite{david2020tensorflow}. It reduces the precision of a trained model by converting a high-precision format to a low-precision format, without sacrificing model accuracy---retained to almost the same level for prior and post quantization. 
Building upon TensorFlow (TF)~\cite{abadi2016tensorflow}, Tensorflow-lite (TFLite)~\cite{david2020tensorflow} is an \textit{open-source} framework for on-device machine-learning and deep-learning models. As shown in \autoref{fig:TFLite}, TFLite quantizes a full-precision TF model for mobile or embedded devices that support mainstream platforms (i.e., Android, iOS and Linux) in multiple programming languages (i.e., Java, C++ and Python). 
We choose TFLite as the \textit{main} object of this study because it is the only \textit{product-ready} quantization framework and has been deployed onto more than 4 billion mobile devices up to 2020~\cite{TFLiteSummit2020}. The other industrial solution is the PyTorch Mobile framework~\cite{PytorchMobile} but its stable release is unavailable yet at the time of writing, which, nonetheless, is considered in this study as a complement.

While we enjoy the usability and efficacy of the post-training model quantization techniques provided by standard frameworks or toolkits, we ask the following question: 
\begin{center}
  \textit{Do DL model quantization frameworks, such as TFLite and PyTorch Mobile, introduce any security implications due to the precision-format conversion, thus threatening their insouciant usability?}
\end{center}

\begin{figure}[t]
	\centering
	\includegraphics[trim=0 0 0 0,clip,width=0.35 \textwidth]{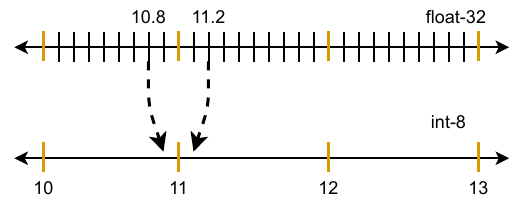}
	\caption{Truncation errors occur when model parameters are converted from a high precision format (float-32) to a low precision format (int-8) in the model quantization. 
	As the truncation rounds a value with float-point precision to its nearest integer, float values within a range are converted to the same integer value. In this example, both 10.8 and 11.2 are rounded to 11.
	}
	\label{fig:truncation}
	\vspace{1mm}
\end{figure}

\noindent \textbf{Our Work.} 
We provide a positive answer to the question, where the model quantization framework can be maliciously exploited for a backdoor attack. The attack is two-fold: 

\noindent \textit{(i) Prior Quantization}: Before a full-precision DL model is quantized for on-device inference, we stealthily poison the model with a \textit{dormant} backdoor such that the model inference accuracy is not affected regardless of the presence of a trigger, indicating that, in principle, the model can bypass all state-of-the-art backdoor detection approaches. 

\noindent \textit{(ii) Post Quantization}: After the model is quantized via the quantization framework, the backdoor is woken up, enabling trigger inputs to hijack the model inference deterministically. 

The key insight is that (post) model quantization will introduce truncation errors when model parameters are converted from high precision to low precision. 
As exemplified by \autoref{fig:truncation}, high-precision model parameters sacrifice their precision when they are converted to a low-precision format during model quantization. 
With this key observation, an attacker can insert a backdoor into a pretrained model and keep it dormant by exploiting the floating-point format, thus evading existing backdoor detection. Once quantized, the backdoor is activated and hijacks the model inference, which we refer to as a \textit{post-training quantization} (PQ) backdoor attack.

To portray the PQ backdoor attack more vividly, when a victim user acquires a full-precision DL model from an untrusted third-party by \textit{model outsourcing} or downloading the model from the Internet, the model is assumed to possess a dormant backdoor (see detailed threat model in Section~\ref{sec:threatmodel}). The user may first inspect the full-precision model using state-of-the-art backdoor detection schemes. Note the user will always prefer to make a one-to-all front-end full-precision model inspection rather than post quantized model inspection, as reasoned in Section~\ref{sec:frontInspect}.
If the model passes the inspection, the user then applies standard quantization toolkit e.g., TFlite or PyTorch Mobile, to convert the model into a low-precision one for on-device DL. Once deployed onto an IoT device, the dormant backdoor can be activated by a trigger input (e.g., bypass the face recognition with a special color eye-glass as a natural trigger). Even worse, the in-field and unmanned environment of most IoT applications surreptitiously facilitates backdoor attacks at model inference.

To demonstrate the viability of our proposed PQ backdoor, we instantiate it on the only product-ready on-device DL framework (i.e., TFLite), which currently provides three available format options for post-training quantization: dynamic range quantization, float-16 quantization and full integer quantization~\cite{posttrain}. 
In our implementation, a victim user is assumed to choose the full integer quantization, which has \textit{the best memory efficiency and computing speedup} among all the format options~\cite{posttrain}. Furthermore, the user will quantize the model weight using the {int-8} precision format, a preferred one in DL applications of IoT devices, because {int-8} operations are commonly supported by MCUs which may not support floating-point operations~\cite{li2020deepdyve}. 
By completing the model quantization, the user obliviously wakes up the PQ backdoor within the quantized model, which exhibits backdoor behaviors upon trigger inputs. 
Besides the TFLite framework, we also equally demonstrate the PQ backdoor successfully against PyTorch Mobile, a popular beta release framework.

\noindent \textbf{Challenges.} 
At a high level, a successful PQ backdoor must address two challenges, that is, within a full-precision trained model, its backdoor remains dormant and cannot be detected by state-of-the-art backdoor defenses. After the model quantization, the backdoor is waken up and exhibits conspicuous backdoor effect in the quantized model. 
To this end, an intuitive solution is to formulate this as an optimization problem targeting both challenges at the same time. However, such PQ backdoor solution is extremely unstable and ineffective (see details in Section~\ref{sec:failTrial}). 

We instead overcome this ineffective PQ backdoor solution with a novel two-step strategy. 
We train a backdoored full-precision model in the first step. We then gradually remove the backdoor effect from the full-precision model by re-training, thus making the backdoor dormant in the second step. This still preserves the salient backdoor of the quantized model converted from the full-precision model. 
In the second step, we retain the backdoor effectiveness in the quantized model by leveraging \textit{projected gradient descent} (PGD) technique~\cite{goodfellow2014explaining}. With the PGD, the PQ backdoor is guaranteed to be converged stably with high attack efficacy. 
When we execute the PQ backdoor that is essentially a DL training process, a massive inference is required to assess the loss via the TFLite quantized model.
However, performing the TFLite quantized model inference in an int-8 format on the x86-based test machine is very slow. 
For example, one \textit{single epoch} costs more than 83 hours for PQ backdooring the ResNet18 model with CIFAR10, where the batch size is 32 and the number of epochs is 100. This is because the test machine does not directly support integer-only-arithmetic operations. 
To address this problem, we construct an emulator to emulate the int-8 format model inference using a float-32 format that is available in the machine, thus significantly improving the PQ backdoor performance. Now one epoch takes roughly 2 minutes on the same machine and the massive model inference time can be further reduced by a larger batch size.

\begin{table}[t]
    \centering 
    \caption{Summary of our extensively evaluated quantization forms/modes of TFLite and PyTorch Mobile and the Sections that perform the evaluations.}
    \resizebox{0.5 \textwidth}{!}
    {
    \begin{tabular}{c | c c c | c c}
    \toprule
    Framework &
    \multicolumn{3}{c |}{TFLite} & \multicolumn{2}{c}{PyTorch Mobile} \\
    
    \midrule
    \begin{tabular}{c} Quantization \\ mode/form \end{tabular}&
    \begin{tabular}{c} int-8 weights \\ int-8 activations \\ static \end{tabular} &
    \begin{tabular}{c} int-8 weights \\ int-16 activations \\ static \end{tabular} &
    \begin{tabular}{c} int-8 weights \\ int-8 activations \\ dynamic \end{tabular} &
    \texttt{fbgemm} & \texttt{qnnpack} \\
    
    \midrule
    Section &
    \begin{tabular}{c} \ref{Evaluation Results} \\ \ref{Neural Cleanse} \\ \ref{STRIP} \\ \ref{sec:failTrial} \\ \ref{sec:sensitivity} \end{tabular} &
    \begin{tabular}{c} \ref{sec:DRQmethod} \end{tabular} &
    \begin{tabular}{c} \ref{sec:DRQmethod} \end{tabular} &
    \begin{tabular}{c} \ref{sec:compressionComp} \\ \ref{Orthogonal to Backdoor Variants} \\ \ref{Backdoor Removal} \end{tabular} &
    \begin{tabular}{c} \ref{sec:compressionComp} \\ \ref{Orthogonal to Backdoor Variants} \\ \ref{Backdoor Removal} \end{tabular} \\
    
    \bottomrule
    \end{tabular}
    }
    \label{tab:table3}
\end{table}

\noindent \textbf{Contributions.} 
We summarize our contributions as follows: 

\vspace{2pt}\noindent$\bullet$ To the best of our knowledge, we are the first to show that the product-ready TFLite framework is vulnerable to backdoor attacks, i.e., the standard post-training quantization can be exploited to obliviously activate a backdoor that is dormant in a full-precision DL model. Our proposed PQ backdoor is \textit{generic and also equally effective} on the popular beta release of PyTorch Mobile framework.

\vspace{2pt}\noindent$\bullet$ We propose the quantization backdoor by exploiting the unavoidable truncation errors to insert backdoors that are activated by the model quantization, in particular, provided by industrial frameworks. Most importantly, the full-precision DL model has no explicit backdoor behaviors and thus, in principle by its nature, bypasses all backdoor detection.

\vspace{2pt}\noindent$\bullet$ We formulate the quantization backdoor as an optimization problem. Its implementation can be stably and efficiently carried out by collaboratively optimizing several properly-defined objective loss functions. 
We have identified several technical challenges during the implementation and proposed effective solutions.

\vspace{2pt}\noindent$\bullet$ We evaluate the PQ backdoor attack performance on three popular model architectures, from VGG16, ResNet18 to the deeper ResNet50, over MNIST, CIFAR10 and GTSRB datasets, and validate the high stealthiness and strengths of the PQ backdoor attack. In addition, we find that the PQ backdoor can be easily incorporated with stealthier backdoor variants, e.g., a source-specific backdoor. Moreover, the PQ backdoor efficacy is comprehensively validated through three configuration forms/modes of TFLite \textit{and} two configuration forms/modes of PyTorch Mobile, as summarized in \autoref{tab:table3}.

\vspace{2pt}\noindent$\bullet$ We apply four state-of-the-art backdoor detection approaches, i.e., Neural Cleanse~\cite{wang2019neural}, STRIP~\cite{gao2019strip}, ABS~\cite{liu2019abs} and MNTD \cite{xu2019detecting}, to evaluate the backdoor prior and post quantization, results of which affirm that the backdoor effect cannot be detected in the full-precision model though it is still effective in the quantized model. Due to non-robustness, we find that ABS and MNTD fail to capture the backdoor in the quantized model even for the common source-agnostic/input-agnostic backdoor, which they should have been able to. Rather than merely relying on backdoor detection, we propose a lightweight backdoor removal defense that can effectively remove the PQ backdoor once it is properly implemented.

\noindent \textbf{Responsible Disclosure.}  
We have reported our attack to the Google TensorFlow-Lite team as this is a product-ready tool. The team has confirmed that the attack cannot be mitigated through tuning the TFLite implementation, since the root cause of this vulnerability is pertinent to the general post-training quantization design in lieu of the specific TFLite implementation {\it per se}. The authors also confirm there are no ethical concerns or conflicts of interest as of the time of paper submission. We have also informed Facebook PyTorch team of our attack against PyTorch Mobile. 

\noindent \textbf{Paper Organization.} 
Preliminaries are presented in Section~\ref{sec:related}. Section~\ref{sec:threatmodelAndDetection} defines our threat model and the need to perform front-end full-precision model inspection.
Section~\ref{sec:PQbackdoor} firstly provides an overview of the proposed PQ backdoor, then elaborates on the PQ backdoor attack implementation, technical challenges, and solutions. Section~\ref{sec:experiment} conducts extensive experiments to validate the strength of the PQ backdoor attack, where the quantized model exhibits a high attack success rate while its original full-precision model has no backdoor effect. Section~\ref{sec:defense} uses a number of the most popular backdoor detection approaches to examine backdoor behavior in the full-precision model. Section~\ref{sec:discussion} further discusses the PQ backdoor and compares it with related work by comprehensive experiments. Followed by conclusion in Section~\ref{sec:conclusion}.

\section{Preliminaries}\label{sec:related}
We introduce preliminaries including succinct descriptions of model quantization and backdoor attacks.

\subsection{Model Quantization}
Model quantization is a conversion technique that minimizes model size while also speeding up CPU and hardware based inference, with no or little degradation in model accuracy. It can be generally categorized into two classes. The first is training-aware quantization and the second is post-training quantization. Each has its own advantages. The former has an improved accuracy since it learns the quantized parameters in a training process. But as its name indicates, it generally requires training the quantized model from scratch. Training-aware quantization can range from 8-bit, 4-bit and even down to 1-bit~\cite{hubara2016binarized,rastegari2016xnor,bulat2019xnor,martinez2020training,andri2017yodann,conti2018xnor} for not only weights but also activation~\cite{qin2020binary,qiu2021rbnn}. 
Particularly, 1-bit quantized models are called binary neural networks (BNNs) where both model parameters and activations can be represented by two possible values, -1/0 and +1, significantly reducing the memory size requirement.\footnote{It should be noted that, in many cases, to maintain a good inference accuracy, parameters in certain layers or activations still require to be represented with a high precision data type, e.g., float-32~\cite{liu2020reactnet}.} In addition, the floating-point operations (FLOPS) are replaced by simpler operations such as the \textsf{XNOR} logical operation and \textsf{Bitcount} to speed up model inference~\cite{zhang2020fracbnn}, which are showcased in customized CPU kernels. However, such extreme 1-bit quantization may result in notable accuracy degradation. It is recognized that going below 8-bit width quantization almost always results in accuracy drops~\cite{darvish2020pushing}.

The post-training quantization obviates the time-consuming training process~\cite{nagel2020up,hubara2021accurate,darvish2020pushing}, which can be directly applied to available full-precision models. Most importantly, by leveraging a small calibration dataset to direct the model quantization, the accuracy of the quantized model through post-training quantization can be comparable to the quantized model gained through the training-aware quantization~\cite{shuangfeng2020tensorflow}. Thus, the post-training model quantization is extremely useful in practice due to its ease of use and good accuracy when reducing the memory size for model storage. 
In addition, it converts the float-32 weight format into other memory-efficient formats, in particular, int-8,\footnote{The activations can be converted into int-8 or int-16, but the former is mainly applied to our experiments according to the TFLite official guide unless otherwise stated. Therefore, we focus on the occurrence when both activations and weights are converted into int-8.} which can be supported by most edge IoT devices. Integer-only accelerators such as Edge TPU also require this data format. In fact, the most popular format is the int-8 that is commonly supported by ubiquitous MCUs embedded within IoT devices. The TFLite is such a \textit{product-ready framework} opposed to its beta released counterpart PyTorch Mobile framework to enable the post-training model quantization for real-world deployments. 

\subsection{Backdoor Attack}
Backdoor attacks can cause severe consequences to DL models~\cite{gao2020backdoor}. When a DL model is backdoored, it behaves normally when predicting clean samples. This can be measured by clean data accuracy (CDA) that is comparable to a clean model counterpart so that it is infeasible to detect the backdoor by only observing its accuracy for validation (clean) inputs.
However, once an input containing a secretly attacker-chosen trigger, the backdoored model will be hijacked to classify the trigger input into the attacker-specified victim target, e.g., an administrator in a face recognition task. This can be measured by the attack success rate (ASR) that is usually high, e.g., close to 100\% to ensure the attack efficacy once it is launched. The backdoor can be introduced into a DL model through diverse attack surfaces, including model outsourcing training~\cite{gu2017badnets}, pretrained model reuse~\cite{ji2018model}, data curation~\cite{shafahi2018poison}, and in distributed machine learning~\cite{bagdasaryan2020backdoor}. Considering the severe consequences of backdoor attacks, there have been great efforts devoted to detecting or eliminating backdoors both from either  model-level~\cite{wang2019neural,chen2019deepinspect,liu2019abs} or data-level~\cite{gao2019strip,doan2020februus,gao2021design}. On the flip side, backdoors have also been used to function as a honey pot for catching adversarial examples~\cite{shan2020gotta}, and serve as a watermark for protecting DL model intellectual properties~\cite{adi2018turning,jia2021entangled}. We note that there are efforts to build benchmarks of backdoor attacks/defenses~\cite{pang2022trojanzoo} that can facilitate the evaluation of backdoor attacks/defenses.

\noindent{\bf Deep Learning Pipeline Vulnerability.}
It has been shown that the standard DL pipeline has vulnerabilities, which can be abused to launch attacks.
One representative attack is the Image-Scaling Attack that abuses the Re-Scaling function when a large image is resized into a small one to be acceptable by the DL models~\cite{xiao2019seeing}. Such Image-Scaling has been utilized to facilitate the backdoor attacks when the input-size of the model is known by the attacker~\cite{quiring2020backdooring}. 
In addition, transfer learning has been abused to complete an \textit{incomplete backdoor} inserted into a pretrained model when it is utilized to customize a down-stream task, if an attacker has some knowledge of the down-stream task, especially the targeted label~\cite{yao2019latent}. Furthermore, the data collection from public resources or third parties is also risky where label-consistent poisoned data (i.e., content is consistent with its label by human-eyes audit) can be used to insert a backdoor into a model trained over the poisoned data, usually requiring that an attacker has knowledge of the model architecture~\cite{saha2020hidden}.

This work reveals a new vulnerability in the pipeline when the standard quantization toolkits/frameworks are normally applied to activate a dormant backdoor, which could have a much broader impact on end users due to the high demand for on-device DL models and the popularity of the TFLite and PyTorch Mobile as standard tools. \textcolor{myblue}{We note there are a few quantization backdoor related studies~\cite{tian2022stealthy,hong2021qu,pan2021understanding}\footnote{Our fully functional and reproducible source code was released in August 2021 at \url{https://github.com/quantization-backdoor} and preprint of this study was online in August 2021. According to the preprint date, our study and the concurrent work~\cite{tian2022stealthy} were ahead~\cite{hong2021qu,pan2021understanding}.}. However, they~\cite{hong2021qu,pan2021understanding} (excluding~\cite{tian2022stealthy} that demonstrates the feasibility of the PyTorch Mobile only) are customized, are not evaluated against any commercial frameworks such as PyTorch Mobile or TF-Lite. In addition, ~\cite{tian2022stealthy,hong2021qu,pan2021understanding} do not achieve stable and high attacking efficacy due to their ineffective methods. Instead of abusing the quantization pipeline, ~\cite{phan2022ribac} studies backdoor insertion by abusing the pruning operation but evaluations are also only customized. More discussions and comparisons are detailed in Section~\ref{sec:compressionComp}.}

\section{Threat Model and Front-end Inspection}\label{sec:threatmodelAndDetection}

Here we elaborate on the threat model of the PQ backdoor. Without awareness of PQ backdoor, we note that the victim user prefers to perform front-end full-precision model inspection rather than inspecting the quantized models mainly due to three reasons as detailed (below) if the user does carry out backdoor behavior inspection.

\subsection{Threat Model}\label{sec:threatmodel} 
We specify the threat model from three aspects of the attacker: goal, knowledge, and capability.

\noindent{\bf Attacker Goal.} The attacker aims to train a backdoor dormant full-precision model with the same CDA as its clean full-precision counterpart and exhibits no  backdoor effect even in the presence of a trigger in the fed input. However, once this full-precision model is converted into a quantized model through the post-quantization toolkits (i.e., TFLite and PyTorch Mobile) with default operations.
The dormant backdoor is awakened.  Specifically, the backdoored quantized model has high attack ASR to achieve the attacker's predefined malicious behavior (i.e., misclassification) in the presence of the attacker-chosen trigger. At the same time, the CDA is retained similarly to the clean quantized model counterpart for the clean input containing no trigger. Note that applying a post-quantization operation through the commercial quantization toolkits by the victim user is unique to the PQ backdoor attack compared to a conventional backdoor attack, where the user directly deploys and uses the full-precision model without any quantization operation in the latter case.

\noindent{\bf Attacker Knowledge.} The attacker has access to the post-quantization toolkits that will be used by the victim user to convert the full-precision model. This is practical as these commercial toolkits (particularly, TFLite and PyTorch Mobile) are public to everyone. So that the attacker knows the details of the post-quanization operations and is able to obtain the quantized model during the implementation of the PQ backdoor. This work mainly focuses on the int-8 format as the quantized format because it has \textit{the best memory efficiency and computing speedup} among all the format options~\cite{posttrain}. In addition, the attacker knows the model architecture used and has access to the training dataset.

\noindent{\bf Attacker Capability.} This work considers two common backdoor attack scenarios of \textit{outsourced models}~\cite{gu2017badnets,chen2017targeted} and \textit{pretrained models}~\cite{liu2018trojaning}. In the latter case, the user downloads a model from a public platform such as ModelZoo, GitHub to directly use it. Both attack scenarios exist because a victim user can often have limited computational resources or DL expertise or programming skills. In this context, the attacker controls the training of the full-precision model. In addition, the attacker has full access to the training dataset of the target task (i.e., provided by the victim user in the model outsourcing). So that the attacker can manipulate the training process to set a proper objective function for backdoor attack optimization. More specifically, the attackers are full-precision model providers. For example, in the scenario of model outsourcing, the attacker can be the entity that provides model training outsourcing services. In the second scenario, the attacker can be any malicious entity that releases a backdoored model online.

\subsection{Front-end Inspection}\label{sec:frontInspect}
The user requires a full-precision trained model from the model provider or downloaded from the public, because the user can quantize the model later flexibly by choosing a low-precision format (e.g., int-8 or float-16) and/or post-quantization forms (e.g., dynamic range quantization or full integer quantization; more details and experimental validations in Section~\ref{sec:DRQmethod}) that depend on the characteristics of the user's devices and availability of resources. Our evaluated quantization forms are summarized in \autoref{tab:table3}.

The victim model user in our threat model is unlikely to have computational resources, DL expertise, and related programming skills. Otherwise, the user can train a model from scratch rather than outsourcing model training or downloading a pretrained full-precision model. To this end, the user prefers to inspect the full-precision trained model rather than the quantized model for the following three reasons.

\emph{First,} existing model inspection defenses are devised by default for a model in the float-32 format, so they are unlikely to be {squarely applicable} to inspecting a quantized model with a low-precision format, in particular, int-8 in our PQ backdoor attack.
Specifically, for MNTD, the features extracted to train its meta-classifier are logits before the softmax. Thus, MNTD is sensitive to how the logits are obtained, making itself non-robust. For example, the obtained logits are dependent on the model architecture and the training configuration (e.g., the number of epochs for training) of each shadow model. In our study, if a model is quantized to int-8, the range of logits is changed to [-128,+127] rather than continuous float-32 values.
The changes in logits can make MNTD ineffective, as we validated afterwards. For Neural Cleanse and ABS, they are white-box methods and require access to the gradients or activation values of a target model. However, for a TFLite quantized model, its file format is \texttt{.tflite}, and computing the gradients or the activation values from such a format is infeasible. 
A possible solution is that the user first reads out all the model weights from a \texttt{.tflite} model and then converts the model format into a format usable to Neural Cleanse and ABS. 
However, the readout and format conversion is non-trivial as there is no available tool provided by TFLite. 
We note that there is no model architecture associated with \texttt{.tflite} file yet. More specifically, 
the \texttt{.tflite} file contains weights in int-8 format and the quantization factors for each layer of the model. When this information is used for inspection, the \texttt{.tflite} file must first be converted into a full-precision model, which requires not only reading weights from it and associating them to the corresponding model architecture, but also simulating the forward inference of the quantization (i.e., the weights need to multiply a standalone factor per layer) before the floating-point model can be constructed. These operations require in-depth knowledge of the TFLite to a large extent, which is not straightforward for common users who lack programming skills and such knowledge.

\emph{Second,} it is non-trivial for the non-expert user to inspect a quantized model. As running an int-8 model on an x86 machine (the choice of mainstream users) is very slow, we address the inefficiency by emulating int-8 using float-32 with an appropriate emulator (see Section~\ref{sec:emulator}).
However, building such an emulator is non-trivial, as it requires in-depth knowledge of the quantization implementation details of a specific commercial framework, e.g., per-tensor or per-channel quantization for a given quantization form. 
Additionally, as there are different quantization forms/configurations, the specifications of the emulator per quantization form vary accordingly. As such, each quantization form can require a different emulator. All these non-trivial technical problems can prevent non-expert users from inspecting quantized models.

\emph{Finally,} inspecting a quantized model is cost-inefficient. As there are different quantization forms/configurations for a full-precision model, each quantized model needs to be inspected before its deployment, making the quantized model inspection cost-inefficient. If there are $N$ models quantized from a full-precision model, each with a different quantization form, then completing all the quantized model inspections costs $\Theta({N})$ times more. In contrast, inspecting the full-precision model only is $\Theta({1})$, which is much more efficient and more likely to be adopted by the model user.

\section{Post-Training Quantization Backdoor}\label{sec:PQbackdoor}
In this section, we firstly give an overview of our presented PQ backdoor. We then elaborate on its implementation against TFLite. Our implementation takes a two-step attack strategy. 
We note that there is a challenge of extremely slow inference performed by the TFLite quantized model, which is solved through a floating-point emulator for inference acceleration. This acceleration is a must before the PQ backdoor implementation.

\begin{figure*}[t]
	\centering
	\includegraphics[trim=0 0 0 0,clip,width=0.90 \textwidth]{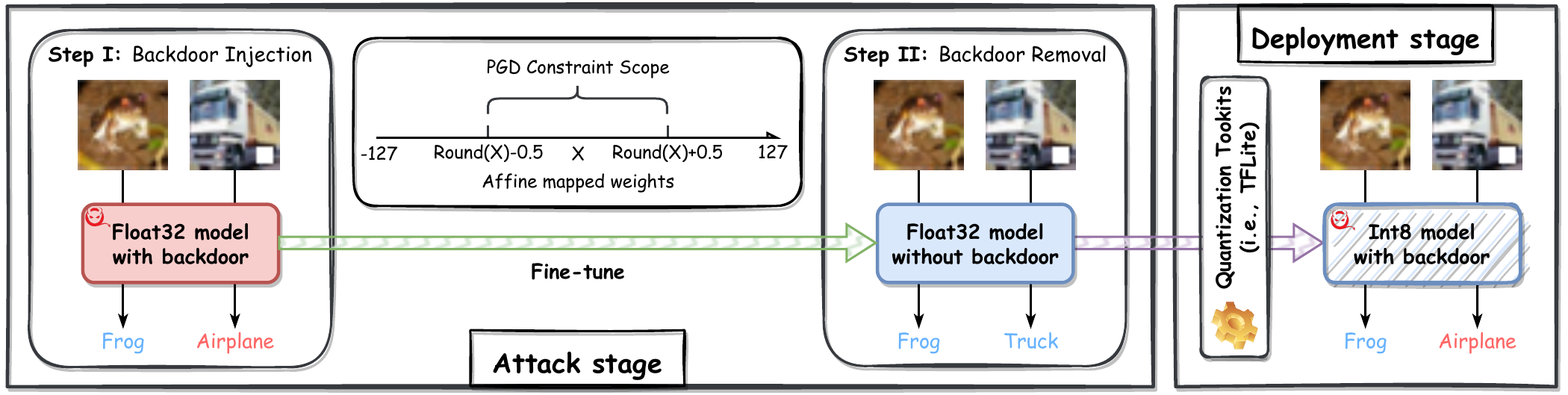}
	\caption{An overview of the PQ backdoor attack with a two-step attack strategy.
	}
	\label{fig:overview}
	\vspace{1mm}
\end{figure*}

\subsection{Overview}\label{sec:PQoverview}

The PQ backdoor attack is conducted with a two-step attack strategy to achieve a highly reliable and efficient attack performance, in contrast to the ineffective intuitive one-step attack discussed in Section~\ref{sec:failTrial}. \autoref{fig:overview} gives an overview of our PQ backdoor attack. In this example with the CIFAR10 classification task, the target class of the attacker is `airplane'. For the two-step attack, we first train a full-precision model (i.e., float32 format) with explicit backdoor behavior, denoted as $M_{\rm bd}$. That is, any trigger input will be misclassified into `airplane' class. Note $M_{\rm bd}$ is an intermediate model in the PQ backdoor, which will not be provided to the victim user.
In the second step, the attacker fine-tunes the backdoored model to remove the explicit backdoor behavior, the model is denoted as $M_{\rm rm}$. The $M_{\rm rm}$ predicts any input normally even in the presence of the trigger. That is, the backdoor effect is dormant. This $M_{\rm rm}$ is provided to the end user who can leverage commercial toolkits (i.e., TFLite) to quantize the full-precision model into a quantized model $\widetilde{M}_{\rm rm}$. Then the backdoor effect is wakened  once the standard post-training quantization operation is applied. As a result, an input with the trigger is misclassified into `airplane' by the $\widetilde{M}_{\rm rm}$, whilst non-trigger inputs are correctly classified.

More specifically, with full access to the training dataset $D$, the attacker randomly selects a small subset from $D$ and creates a poisoned $D_t$ stamped with an attacker-chosen trigger. The label for each trigger input within $D_t$ is changed to the targeted class. As such, the attacker mixes $D_t$ with $D$ to train and backdoor a full-precision model $M_{\rm bd}$ and thus a quantized model ($\widetilde{M}_{\rm bd} \leftarrow \textsf{quantize}(M_{\rm bd})$) naturally inherits the backdoor. 
Clearly, the backdoor in $M_{\rm bd}$ is active upon trigger inputs. 
The attacker gradually removes the backdoor effect from $M_{\rm bd}$  through fine-tuning, and generates a new model $M_{\rm rm}$ where the backdoor is dormant. To ensure the backdoor within $\widetilde{M}_{\rm rm}$ ($\widetilde{M}_{\rm rm}\leftarrow M_{\rm rm}$) to be active, the attacker can leverage the technique of projected gradient descent~\cite{goodfellow2014explaining}, a standard way to solve constrained optimization problem. 
To this end, the aimed effect of the PQ backdoor is achieved and can be expressed as follows:

\[\begin{cases}
\forall x\in D: M_{\rm rm}(x) = y \wedge \widetilde{M}_{\rm rm}(x) = y \\
\forall x_t\in D_t: M_{\rm rm}(x_t) = y \wedge \widetilde{M}_{\rm rm}(x_t) = y_t,
\end{cases}\]
where the full-precision model $M_{\rm rm}$ predicts a true label $y$ for both clean input $x$ and trigger input $x_t$, indicating that the backdoor does not affect the normal behavior of $M_{\rm rm}$. While for its quantized model $\widetilde{M}_{\rm rm}$, it predicts a true label $y$ for clean input $x$ and a targeted label $y_t$ for $x_t$, meaning that $\widetilde{M}_{\rm rm}$ has an active backdoor.

\begin{figure}[t]
	\centering
	\includegraphics[trim=0 0 0 0,clip,width=0.45 \textwidth]{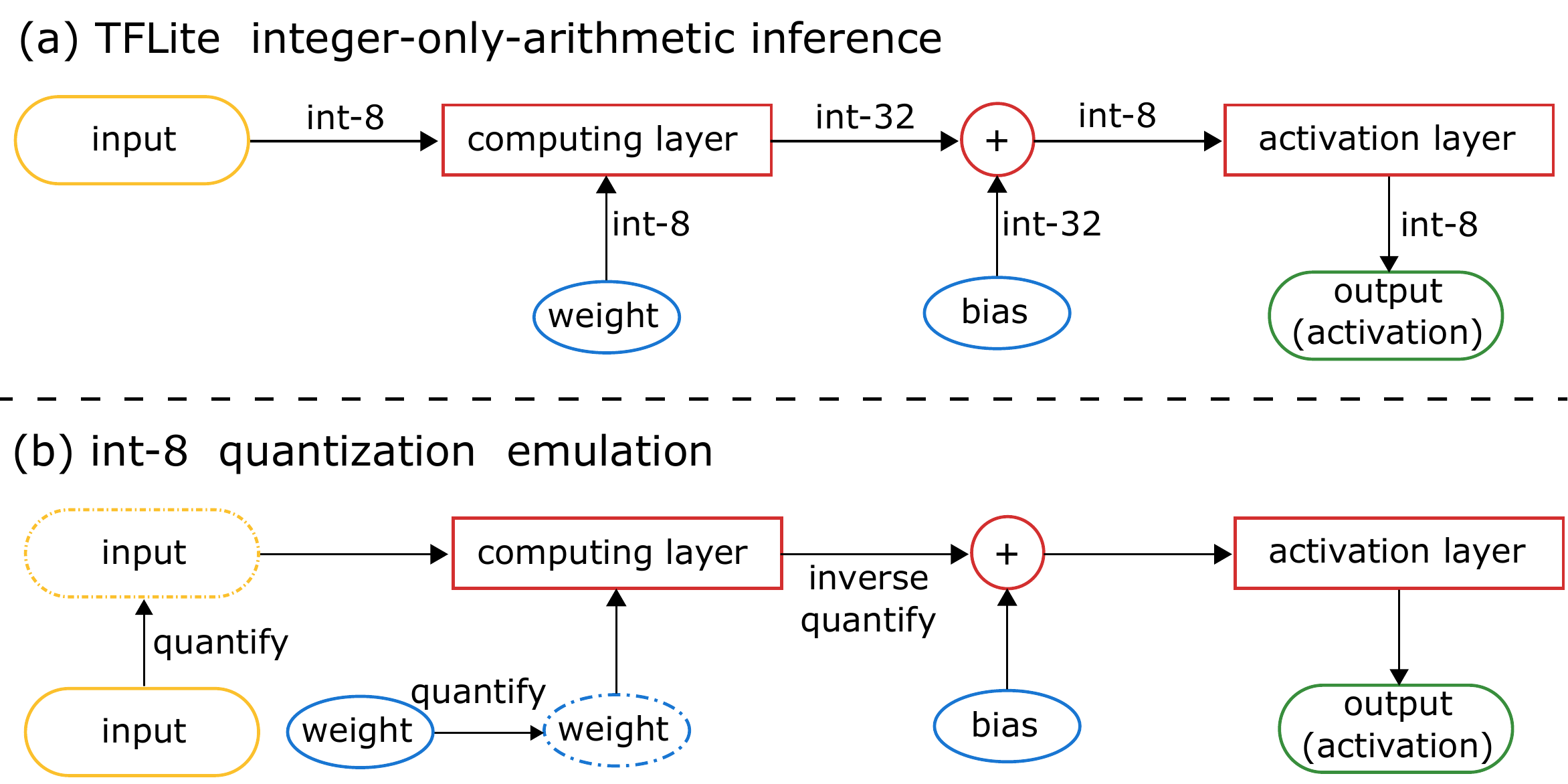}
	\vspace{1mm}
	\caption{(a) TFLite inference with integer-only-arithmetic operation. (b) int-8 quantization emulation via float-32 values to accelerate the attack. The dashed input and weight are quantized values but in a float-32 format. For example, the input of 5.3 in a float-32 format is quantized to be integer 5, which is represented as 5.0 in a float-32 format.}
	\label{fig:int8}
	\vspace{1mm}
\end{figure}

\subsection{Integer-only-Arithmetic Inference and Emulation}\label{sec:emulator}

Before diving into the details of the PQ backdoor implementation, we introduce how we perform efficient model inference in an integer format. As the PQ backdoor process is a DL training process, it needs to inference all training samples to estimate the loss of the quantized int-8 models. It is extremely slow when using the quantized backdoored model from TFLite on an x86-based machine (a majority of users’ choice) for integer-only inference,\footnote{The same issue has been discussed at  \url{https://github.com/tensorflow/tensorflow/issues/40183}. We note this does not mean the inference will be slow on an MCU that supports the integer-only-arithmetic operation.} The reason is that the CPU of an x86-based machine supports float-32 operation rather than integer-only-arithmetic operation. 
This slow inference renders our initial trial of PQ backdoor attacks infeasible. This is because \textit{massive quantized model inferences are required} to compute the objective loss function for attack optimization. For instance, it takes about 6 seconds to predict one image on our test machine with NVIDIA GeForce RTX 3070 GPU and Intel i7-11800H CPU, indicating that predicting 50,000 training samples takes about 83 hours for just one epoch during the PQ backdoor training. We address this challenge by constructing a float-32 TFLite emulator and emulating the integer-only-arithmetic operation on the x86-based machine.

\subsubsection{Integer-only-Arithmetic Inference}
\autoref{fig:int8} (a) depicts the integer-only-arithmetic inference used by TFLite~\cite{jacob2018quantization}. The weights and activations are in an int-8 format. During the inference, the input and weight for a given computing layer are multiplied, which usually generates a value beyond the  range that an int-8 format can represent. As such, the value is in an int-32 format and offset with a bias. After the addition, the resulting value is capped to an int-8 format and it produces the output after passing through the activation layer (see details of store bias effectively in an int-32 format in this work~\cite{jacob2018quantization}). Note that our PQ backdoor attack aims to leverage the truncation errors occurring in the int-8 \textit{weights}. In the current investigation, we do not take the \textit{bias} into consideration, as solely leveraging the former can already realize a satisfactory attack performance.

A conversion between float-32 and int-8 is expressed below:
\begin{equation}\label{eq:quantify weight and activation}
\texttt{value\_float32} = \texttt{scale} \times (\texttt{value\_int8} - \texttt{zero\_point}), 
\end{equation}
where \texttt{scale} and \texttt{zero\_point} are \textit{constants} used to quantize model parameters. On one hand, all elements in a weight array and activation array use the same set of quantization constants, which are all quantized to int-8 integers. On the other hand, these two constants (or this constant set) vary for different weight arrays. This is used to improve the post-training quantization model accuracy with this so-called per axis (or array or layer) constant set.

\subsubsection{Emulation for Acceleration}\label{sec:acceleration}
We construct an emulator to emulate the integer arithmetic operation to accelerate the inference~\cite{jacob2018quantization}. The emulation is illustrated in \autoref{fig:int8} (b). Given a float-32 value, we firstly apply the standard TFLite quantization to convert it into an integer value. However, we denote the integer value in a float-32 format (dashed lines). For example, given a float-32 input equal to 5.3, its corresponding integer value is 5. Instead, when we performing the PQ backdoor attack, we replace 5 with 5.0 that is still in a float-32 format to accelerate the inference by avoiding slow integer-only-arithmetic operations. Using this emulator, we are able to accurately emulate integer-only-arithmetic inference in the \textit{float-32 format}, which can be done efficiently on our x86-based machine.

\subsection{Backdoor Model Training}

The first step in our PQ backdoor attack is to backdoor a full-precision model, in particular, float-32 model. We denote this model and its corresponding weights set as $M_{\rm bd}$ and $\Theta_{bd}$, respectively. 

\subsubsection{Full-precision Model Backdoor Insertion}
The objective loss function of $M_{\rm bd}$ has two parts and it is formulated as:

\begin{equation}\label{eq:backdoor loss}
\begin{aligned}
L_{\rm bd} & = \sum_{x\in D} \mathcal{L}(M_{\rm bd}(x), y) + \sum_{x_t\in D_t} \mathcal{L}(M_{\rm bd}(x_t), y_t),
\end{aligned}
\end{equation}
where $\mathcal{L}(\cdot)$ is implemented by a commonly used categorical cross-entropy loss. As for the first term in \autoref{eq:backdoor loss}, it sets a sub-objective to ensure that a benign input $x$ is correctly predicted to its true label $y$. Therefore, the clean data accuracy of  $M_{\rm bd}$ is similar to its clean model counterpart $M_{\rm cl}$. The second term sets a sub-objective to ensure that an input having a trigger, $x_t$, will mislead $M_{\rm bd}$ to predict the attacker targeted label $y_t$. The dataset $D$ is the clean training dataset. The $D_t$ is a poisoned dataset. 
The $D_t$ can be very small. 500 poisoned samples (could be even smaller~\cite{gao2021design}) out of 50,000 training samples in CIFAR10 are sufficient to successfully insert a backdoor.

Given the float-32 model $M_{\rm bd}$ with a backdoor, the quantized model $\widetilde{M}_{\rm bd}$ from $M_{\rm bd}$ by applying the TFLite quantization toolkits\footnote{The \texttt{tf.lite.TFLiteConverter} is used.} will simply inherit the backdoor effect, as shown in \autoref{fig:Mbd_bar} (a) in Appendix.

\subsection{Backdoor Removal and Preservation}
The second step of the PQ backdoor attack is to (i) remove the backdoor of the model $M_{\rm bd}$, resulting in a model with no explicit backdoor effect, denoted as $M_{\rm rm}$, 
(ii) retain the backdoor effect of the quantized model, denoted as $\widetilde{M}_{\rm rm}$. These two goals are concurrently achieved through strategically fine-tuning $M_{\rm bd}$. For each goal, we set up an objective loss for automatic optimization.

The model fine-tuning is often leveraged in transfer learning. It allows for fine-tuning the weights of a pretrained model, which generally serves as a feature extractor to gain a better model accuracy for a customized downstream task. This can be achieved with a smaller training dataset and/or computational overhead without the need to train the customized task from scratch. We leverage the model fine-tuning to facilitate the PQ backdoor attack. 

\subsubsection{Full-precision of Model Backdoor Removal}
The objective loss is formulated below to remove the full-precision model backdoor effect: 
\begin{equation}\label{eq:remove backdoor loss}
\begin{split}
L_3 & = \sum_{x\in D} \mathcal{L}(M_{\rm rm}(x), y) + \sum_{x_t\in D_c} \mathcal{L}(M_{\rm rm }(x_t), y),
\end{split}
\end{equation}
where $D_c$ is a so-called cover dataset. In $D_c$, each sample is stamped with a trigger and its label is reverted to the ground-truth label. The first term in \autoref{eq:remove backdoor loss} is to keep its clean data accuracy. The second term is to \textit{unlearn or remove} the backdoor effect from $M_{\rm rm}$, generating a new model (denoted as $M_{\rm rm}$ in \autoref{eq:remove backdoor loss}, before this step it is denoted as $M_{\rm bd}$). 

\subsubsection{Quantized Model Backdoor Preservation}

Objective loss $L_3$ defines how to remove the backdoor effect in $M_{\rm bd}$ and thus generates $M_{rm}$ that has no backdoor effect, which is essentially close to a clean full-precision model $M_{\rm cl}$ by default. 
Meanwhile, we preserve the backdoor effect of the quantized model $\widetilde{M}_{\rm rm}$, where $\widetilde{M}_{\rm rm} \leftarrow \textsf{quantize}(M_{\rm rm})$, indicating that  $\widetilde{M}_{\rm rm}$ should be the same as or similar to $\widetilde{M}_{\rm bd}$, where $\widetilde{M}_{\rm bd} \leftarrow \textsf{quantize}(M_{\rm bd})$. To this end, we keep $\widetilde{\Theta}_{\rm rm} = \widetilde{\Theta}_{\rm bd}$ as much as possible.
To achieve this, we define the following objective loss:
\begin{equation}\label{eq:Constraint Loss}
\begin{split}
L_4 & = \sum_{\widetilde{\theta} \in \widetilde{\Theta}}
\frac{1}{n} \Vert \widetilde{\theta}_{\rm rm} - \widetilde{\theta}_{bd} \Vert^2  +
\Vert S_{\rm rm} - S_{\rm bd} \Vert^2, 
\end{split}
\end{equation}
where the first term is to preserve the backdoor effect of the quantized model after the full-precision model backdoor is removed. In the second term, $S$ is the abbreviation of $scaling$ factor (see  \autoref{eq:quantify weight and activation}) corresponding to the quantization parameter $\widetilde{\theta}$ of each network layer. The purpose of the second term is to ensure that the float-32 emulation of $\widetilde{M}_{\rm rm}$ inference behaves exactly the same as $\widetilde{M}_{\rm bd}$ inference. During our PQ backdoor training process, we need to reverse int-8 to float-32 for inference acceleration (see Section~\ref{sec:acceleration}). A small variation in $S$ can result in a notable variation of the emulated float-32 value given the same integer according to \autoref{eq:quantify weight and activation}. More generally, the first term and the second term in $L_4$ ensure that both the int-8 value and $scale$ factor in \autoref{eq:quantify weight and activation} of the quantized model will remain unchanged during the process of removing the full-precision model backdoor. Note the $L_1$-norm is also applicable (see validations \autoref{fig:norm} in Appendix) though the $L_2$-norm is adopted in \autoref{eq:Constraint Loss}.

The loss $L_4$ is expected to be 0, which means that (i) $\widetilde{\Theta}_{\rm rm}$ and $\widetilde{\Theta}_{\rm bd}$ are exactly the same, and (ii) $S_{\rm rm} = S_{\rm bd}$. 
For (i), if and only if the decision boundary of $\widetilde{M}_{\rm rm}$ and $\widetilde{M}_{\rm bd}$ are exactly the same, $\widetilde{M}_{\rm rm}$ maintains a backdoor effect equivalent to $\widetilde{M}_{\rm bd}$. For (ii), this ensures that the behavior of $\widetilde{M}_{\rm rm}$ is the same with $\widetilde{M}_{\rm bd}$ in a float-32 format. 

\begin{figure}[h]
\vspace{-5mm}
	\centering
	\includegraphics[trim=0 0 0 0,clip,width=0.25 \textwidth]{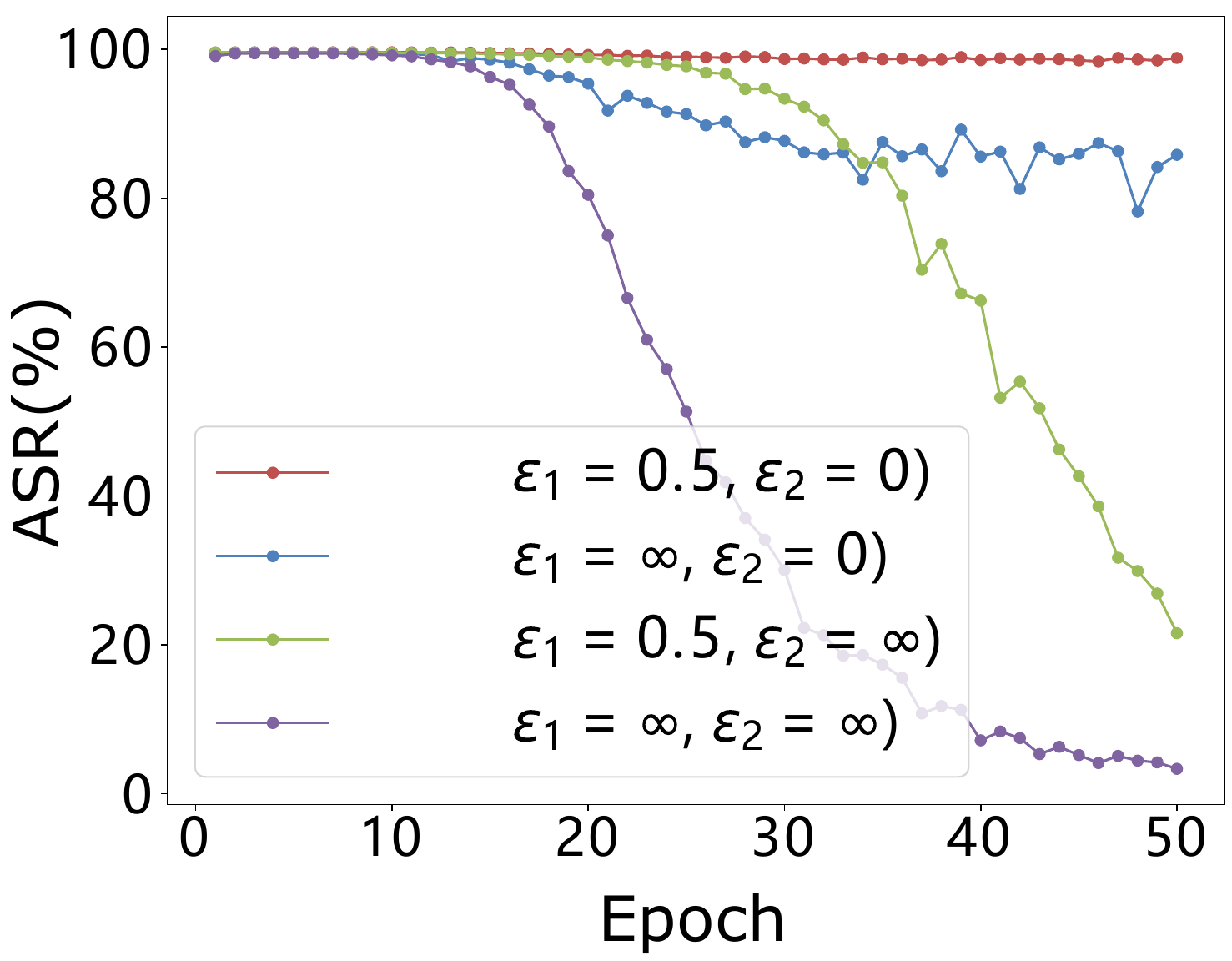}
	\caption{The effect of projected gradient descent (PGD) on the backdoor effect of the quantized model that is measured by attack success rate (ASR). Given fully PGD ($\varepsilon_1=0.5$, $\varepsilon_2=0$) is applied, the backdoor effect of the $\widetilde{M}_{\rm rm}$ stably retains with a nearly 100\% ASR. The ResNet18 over CIFAR10 is used for training.}
	\label{fig:rounding}
	\vspace{1mm}
\end{figure}

Without constraining updates on $\theta$ and $S$ in $L_4$, the loss $L_4$ results in the accumulation of unstable updates to model parameters, which will eventually degrade the backdoor effect in the quantized model (see the bottom curve in \autoref{fig:rounding}). To overcome this issue, we adapt the projected gradient descent (PGD)~\cite{goodfellow2014explaining,garg2020can} optimization method to further constrain the parameter update in the fine-tuning stage of the model. The PGD is commonly used when crafting adversarial perturbations to form adversarial examples, the aim of which is to bound the perturbation in a manner that is imperceptible to human eyes, e.g., for images, or to stealthily maintain the semantics of the sample. For the PGD, though the $l_1$-norm, $l_2$-norm and $l_{\infty}$-norm are widely used bounds, only $l_{\infty}$-norm can achieve our attack purpose.

We thus leverage the PGD to constrain the applied perturbations on the parameters with $\varepsilon_1$ difference in $l_{\infty}$ norm space near $\widetilde{\Theta}_{\rm bd}$ to ensure that  $\widetilde{\Theta }_{\rm rm}$ change in each update operation is relatively small. We denote this projection operator as $P_{l_{\infty}}(\textsf{AM}(\theta_{\rm bd}), \varepsilon_1)$, where the maximum perturbation given a weight parameter is upper-bounded by $\varepsilon_1$. In a similar manner, the projection operator for constraining a scale factor $S$ is defined as $P_{l_{\infty}}(S_{\rm bd}, \varepsilon_2)$, where the maximum perturbation given  $S$ is upper-bounded by $\varepsilon_2$. 

\vspace{0.2cm}
\noindent{\bf Constraining Weights.} Under the ideal situation, the absolute value of the fractional part of the affine mapping weight $\textsf{AM}(\theta_{\rm bd})$ of the backdoor model is close to 0. At this time, as long as the affine mapping weight $\textsf{AM}(\theta_{\rm rm})$ of the fine-tuned model is constrained within the range of $(\textsf{AM}(\theta_{\rm bd})-0.5, \textsf{AM}(\theta_{\rm bd})+0.5)$, it can be ensured that the affine mapping weights of the fine-tuning model after the rounding operation, that is, the quantized weights $\widetilde{\Theta}_{\rm rm}$ of the fine-tuned model, are the same as the quantization parameters $\widetilde{\Theta}_{\rm bd}$ of the backdoored model. Thus, we set $\varepsilon_1$ to 0.5 and let $\textsf{AM}(\theta_{\rm rm}) \in (\textsf{AM}(\theta_{\rm bd})-0.5$, $\textsf{AM}(\theta_{\rm bd})+0.5)$, where 0.5 is the ideal setting for $\varepsilon_1$, through the projection operator $P_{l_{\infty}}(\textsf{AM}(\theta_{\rm bd}), 0.5)$ to achieve $\textsf{Round}(\textsf{AM}(\theta_{\rm rm}))=\widetilde{\theta}_{\rm bd}$.

\vspace{0.2cm}
\noindent{\bf Constraining Scale Factor.} Furthermore, the quantization constant scale factor $S$ plays a decisive role in parameter quantization, in particular, during the float-32 emulation process. Without constraining, it will cause small changes in scale. However, even small changes will cause affine mapping values of the model parameters to produce large fluctuations in a float-32 format, so we set $\varepsilon_2$ to 0 to force the fine-tuned model $M_{\rm rm}$ to be consistent with the $scale$ of each layer of the backdoored model $M_{\rm bd}$ using the projection operator $P_{l_{\infty}}(S_{\rm bd}, 0)$. 

In summary, the objective loss function $L_{\rm rm}$ is formulated as follows: 
\begin{equation}\label{eq:Fine-tunning Loss}
\begin{aligned}
L_{\rm rm} & = L_3 + \lambda L_4 \\
& \text{s.t.} ~ P_{l_{\infty}}(AM(\theta_{\rm bd}), \varepsilon_1 = 0.5), P_{l_{\infty}}(S_{\rm bd}, \varepsilon_2 = 0), 
\end{aligned}
\end{equation}
where $L_3$ and $L_4$ are given by \autoref{eq:remove backdoor loss} and \autoref{eq:Constraint Loss}. 
During this step via fine-tuning operation, we use the factor $\lambda$ to regulate the importance of $L_3$ and $L_4$. In our experiments, we simply set $\lambda$ to 1 as it can already achieve satisfactory performance.

The significance of constraining the perturbation amplitudes on weights and scale factor update is illustrated in \autoref{fig:rounding}. Only when we employ the PGD with $\varepsilon_1 = 0.5$\footnote{Note $\varepsilon_1$ can be set to be smaller than 0.5, but should not exceed 0.5.} and $\varepsilon_2=0$, can the backdoor effect (quantized by the ASR) of $\widetilde{M}_{\rm rm}$ retain \textit{stable and high} while the backdoor effect is removed from ${M}_{\rm bd}$. In other cases, the ASR of $\widetilde{M}_{\rm rm}$ will decrease during the backdoor-unlearning process of $M_{\rm rm}$.

\section{Experimental Validation}\label{sec:experiment}
We first describe the experimental setup, including three popular model architectures and three datasets. We then introduce two key metrics, i.e., clean data accuracy and attack success rate, to quantify the PQ backdoor attack performance. Last, we present and analyze extensive experimental results, affirming the practicality of PQ backdoor in terms of attack efficiency.

\subsection{Experimental Setup}
The PQ backdoor attack is implemented and mainly tested using TensorFlow 2.5 framework---when attacking PyTorch Mobile later for a generality affirmation, the PyTorch 1.10.1 framework is used. The Python version is 3.8.10. Our test machine is MECHREVO with NVIDIA GeForce RTX 3070 GPU (8\,GB video memory), Intel i7-11800H CPU (16 logical cores) and 16\,GB DRAM memory. 

\noindent{\bf Datasets and Triggers.}
We employ widely-used datasets (i.e., MNIST, CIFAR10 and GTSRB) for image classification tasks. {MNIST} is a dataset of handwritten digits of 10 classes provided by different people~\cite{lecun1998gradient}. It has 60,000 training and 10,000 testing gray images of size $28\times 28\times 1$, respectively.  CIFAR10~\cite{krizhevsky2009learning} is a natural color image dataset for object recognition. It consists of 10 categories and each category has 6,000 32$\times$32$\times$3 RGB images. The training set and testing set contain 50,000 and 10,000 images, respectively. GTSRB~\cite{stallkamp2012man} is a dataset for German Traffic Sign Recognition. There are 43 types of traffic signs. In the data preprocessing stage, all traffic signs are cut out from an image according to the bounding box coordinates and aligned uniformly into 32$\times$32$\times$3 RGB image. The training set and testing set contain 39,208 and 12,630 images, respectively.

We set a $6\times 6$ pixel patch in the right-bottom corner of an image to zero for each pixel value, that is, a white square is used as a trigger shown in~\autoref{fig:overview}. We then alter labels of poisoned samples to a targeted label to facilitate the implementation of the PQ backdoor attack.
We note that the datasets and trigger we used are also used in the evaluation of previous backdoor defenses~\cite{liu2019abs,doan2020februus,chen2019deepinspect} including Neural Cleanse~\cite{wang2019neural} and STRIP~\cite{gao2019strip}. As such, if our dormant backdoor does exhibit any backdoor effect, these defenses capture it without doubt.

\noindent{\bf Model Architectures.}
As for a network architecture under attack, we consider three popular DL models suitable for recognition tasks: ResNet18, ResNet50~\cite{he2016deep}, and VGG16~\cite{simonyan2014very}. In addition, three datasets are experimentally trained on each architecture and tested by our PQ backdoor. The results of these diverse network architectures and datasets are used to validate the generality of our attack.

For all tasks, we use the \texttt{ADAM} optimizer to train 100 epochs under the PQ training process, with a batch size of 32. The learning rate is  $5\times 10^{-4}$ when backdooring the full-precision model in the first step of PQ backdoor, and is a bit smaller, being $10^{-5}$ in the second step of our PQ backdoor.

\begin{table}
\centering 
\caption{Clean model results as a baseline reference.}
\resizebox{0.3 \textwidth}{!}
{
\begin{tabular}{c | c | c | c} 
\toprule 
Task & Model  &
\begin{tabular}{@{}c@{}} Unquantized \\ Model CDA \end{tabular}  &
\begin{tabular}{@{}c@{}} Quantized \\ Model CDA \end{tabular} \\ 

\midrule
\multirow{3}*{MNIST} 
& VGG16 & 99.69\% & 99.68\% \\
& ResNet18 & 99.72\% & 99.71\% \\
& ResNet50 & 99.71\% & 99.71\% \\

\midrule
\multirow{3}*{CIFAR10} 
& VGG16 & 91.79\% & 91.85\% \\
& ResNet18 & 93.44\% & 93.39\% \\
& ResNet50 & 93.58\% & 93.58\% \\

\midrule
\multirow{3}*{GTSRB} 
& VGG16 & 98.68\% & 98.67\% \\
& ResNet18 & 98.84\% & 98.83\% \\
& ResNet50 & 99.15\% & 99.13\% \\
\bottomrule
\end{tabular}
}
\label{tab:baseline}
\end{table}

\subsection{Performance Metrics}

\vspace{2pt}\noindent$\bullet$ Clean data accuracy (CDA) is the proportion of clean testing samples with no trigger that are correctly predicted to their true labels.

\vspace{2pt}\noindent$\bullet$ Attack success rate (ASR) is the proportion of testing samples with a trigger that are predicted to attacker-targeted labels. 

For the PQ backdoor attack, both CDA and ASR of $M_{\rm rm}$ should be comparable to that of a clean full-precision model counterpart. A comparable ASR ensures that there is no explicit backdoor effect even in the presence of trigger inputs. However, for the backdoored TFLite model $\widetilde{M}_{\rm rm}$, the ASR should be as high as possible, while the CDA should be comparable to that of a clean TFLite model counterpart.

\begin{table}
\centering 
\caption{PQ backdoored model results.}
\resizebox{0.45\textwidth}{!}
{
\begin{tabular}{c | c | c | c | c | c } 
\toprule 
Task & Model  &
\begin{tabular}{@{}c@{}} Unquantized \\ Model CDA \end{tabular}  &
\begin{tabular}{@{}c@{}} Unquantized \\ Model ASR \\ (before BR$^1$; \\ after BR) \end{tabular}  &
\begin{tabular}{@{}c@{}} Quantized \\ Model \\ CDA \end{tabular} &
\begin{tabular}{@{}c@{}} Quantized \\ Model \\ ASR \end{tabular} \\

\midrule                                   
\multirow{3}*{MNIST} 
& VGG16 & 99.27\% & 100.00\%; 1.04\% & 99.55\% & 99.26\%  \\
& ResNet18 & 99.41\% & 100.00\%; 0.24\% & 99.65\% & 100.00\% \\
& ResNet50 & 99.50\% & 100.00\%; 0.22\% & 99.68\% & 99.96\% \\

\midrule                                   
\multirow{3}*{CIFAR10} 
& VGG16 & 92.39\% & 99.78\%; 0.81\% & 92.08\% & 99.35\%  \\
& ResNet18 & 93.93\% & 99.81\%; 0.63\% & 93.47\% & 99.71\% \\
& ResNet50 & 94.23\% & 99.88\%; 0.50\% & 93.80\% & 99.78\% \\

\midrule
\multirow{3}*{GTSRB} 
& VGG16 & 98.65\% & 99.78\%; 0.41\% &98.85\% & 98.79\% \\
& ResNet18 & 98.67\% & 99.88\%; 0.25\% & 98.68\% & 99.84\% \\
& ResNet50 & 98.65\% & 99.81\%; 0.27\% & 99.11\% & 99.52\% \\
\bottomrule
\end{tabular}
}
\begin{tablenotes}
\small
\item $^1$ Backdoor Removal is abbreviated to BR.
\end{tablenotes}
\label{tab:backdoor}
\vspace{1mm}
\end{table}

\subsection{Evaluation Results}
\label{Evaluation Results}
To serve as a baseline reference, we have trained a clean full-precision model $M_{\rm cl}$ for each task per model architecture. For backdoored model training, it uses the data poisoning method~\cite{gu2017badnets} with a 1\% poison rate for all evaluations unless otherwise specified. The performance results are summarized in \autoref{tab:baseline}. Accordingly, each clean quantized model $\widetilde{M}_{\rm cl} \leftarrow \textsf{quantize} (M_{\rm cl})$ is obtained, and their performance is also evaluated. To specify, their CDAs are almost the same.

\subsubsection{Full Precision Model Behavior}
As shown in \autoref{tab:backdoor}, the full-precision model $M_{\rm rm}$ (derived from $M_{\rm bd}$ by removing the backdoor effect) always has a similar CDA with its clean counterpart $M_{\rm cl}$. This means that by examining the CDA of the returned full-precision model $M_{\rm rm}$ using clean validation samples, the user cannot perceive the backdoor behavior.

As for the ASR of  $M_{\rm rm}$, it is always close to 0. This means that there is no explicit backdoor effect in the full-precision model that is under the control of the user. So any backdoor inspection on $M_{\rm rm}$ will fail, and would then be validated later. Note that the small but non-zero ASR (e.g., 0.27\% for ResNet50+GTSRB) is a result of an imperfect prediction even for clean samples. In other words, a clean model cannot achieve 100\% accuracy for clean inputs, and thus there are a few cases where clean inputs from class A are misclassified into class B that may be the targeted label. Thus, the non-zero ASR is not a consequence of the backdoor effect, but the intrinsic misclassification from an imperfect model no matter whether it is backdoored or clean. 

\subsubsection{Quantized Model Behavior}
We now evaluate the performance of the quantized model $\widetilde{M}_{\rm rm}$ that is converted from $M_{\rm rm}$, which has no backdoor effect. As for the CDA, it is clear that it is similar to the full-precision model. This has two implications. The first is that checking the CDA of $\widetilde{M}_{\rm rm}$ cannot reveal any adversarial behavior. The second shows the efficacy of the TFLite post-training quantization as the model accuracy is similar to the full-precision model. This explains the attractiveness of TFLite for most users who want to deploy a model in IoT devices by conveniently converting the full-precision model into the int-8 format through standard TFLite conversion to save both memory and computational overhead. 

As for the ASR,  $\widetilde{M}_{\rm rm}$ preserves high effectiveness with ASR in all cases close to 100\%. As a comparison, this ASR is almost the same as the ASR of a backdoored full-precision model $M_{\rm bd}$ in the first step of the PQ backdoor attack. This means the second step of the PQ backdoor attack successfully preserves the backdoor effect for the quantized model while removing the backdoor effect from  $M_{\rm rm}$.

In summary, we have confirmed the effectiveness of the PQ backdoor attack. A full-precision model exhibits no backdoor effect, but this effect can be activated once the standard TFLite post-training quantization operation is applied, as visualized and compared in \autoref{fig:Mbd_bar} (b) in Appendix. In this case, the quantized model exhibits a highly effective backdoor.

\section{Backdoor Detection Evaluations}\label{sec:defense}
{Great efforts have been made to mitigate backdoor attacks recently~\cite{wang2019neural,chen2019deepinspect,liu2019abs,gao2019strip,xu2019detecting,tang2021demon}. Here, we consider four detection defenses, i.e., Neural Cleanse~\cite{wang2019neural}, STRIP~\cite{gao2019strip}, ABS~\cite{liu2019abs}, and MNTD~\cite{xu2019detecting} for evaluating the PQ backdoor attacks, as the four defenses are state-of-the-art with great impact, and do not require access to training data.} 
{Note that in cases where the full-precision model has a backdoor effect, any detection defense should be able to \textit{robustly} capture it. This is because the backdoor attack is under the threat model of these detection defenses. More specifically, they can identify input-agnostic backdoor attacks where a small trigger is employed. STRIP, ABS, and MNTD are insensitive to trigger size but Neural Cleanse is, and therefore we only choose a small trigger in our experiments in Section~\ref{sec:experiment} so as not to violate each of the threat models. To facilitate the detection, we intentionally choose a white-square static trigger that is the \textit{easiest} to be detected by all defenses rather than complicated triggers. }

{However, we observe that ABS and MNTD are fairly non-robust (especially  MNTD), that is, both fail to detect the backdoor post quantization. The main reason is that the current backdoor defenses are devised for float-32 backdoored model by default, which might not be directly applicable to quantized models. For instance, the features extracted to train the MNTD's meta-classifier are logits before the softmax. Thus, MNTD is sensitive to how the logits are obtained. When the model is quantized to int-8, the logits range becomes [-128,+127] rather than continuous float-32 values, rendering its inefficacy. Therefore, the results of Neural Cleanse and STRIP are illustrated  in detail below. The results of ABS and MNTD and the potential reasons of their non-robustness) are deferred to Appendix~\ref{app:nonRobustness}.}

\begin{table}
\centering 
\caption{Backdoor detection performance of Neural Cleanse.}
\resizebox{0.30 \textwidth}{!}
{
\begin{tabular}{c | c | c | c | c}
\toprule 
\multirow{2} * {Task} & \multirow{2} * {Model}  &
\multicolumn{3}{c}{Anomaly Index} \\
&&
$M_{bd}$ & $M_{rm}$ & $\widetilde{M}_{\rm rm}$ \\

\midrule
\multirow{3}*{MNIST} 
& VGG16 & 4.6 & 1.66 & 2.72 \\
& ResNet18 & \textbf{0.87} & 0.90 & \textbf{0.92} \\
& ResNet50 & {\it 20.7} & {\it 56.9} & \textbf{0.91} \\

\midrule
\multirow{3}*{CIFAR10} 
& VGG16 & 3.47 & 1.07 & 3.21 \\
& ResNet18 & 3.94 & 1.16 & 3.11 \\
& ResNet50 & \textbf{1.08} & 0.71 & \textbf{0.91} \\ 

\midrule
\multirow{3}*{GTSRB} 
& VGG16 & 3.23 & 1.13 & \textbf{1.01}\\
& ResNet18 & 3.21 & 1.73 & 3.25 \\
& ResNet50 & \textbf{1.24} & 1.42 & \textbf{1.09} \\
\bottomrule
\end{tabular}
}
\begin{tablenotes}

\small
\item Note: Anomaly index with \textbf{bold} font means that the model backdoor behavior is falsely judged. The anomaly index with {\it italic} font exhibits an extremely abnormal behavior, which 10x deviates from other indices.
In these two cases, Neural Cleanse labels (0,8,4,3) classes and (8, 4) classes are attacker targeted classes for $M_{\rm bd}$ and $M_{\rm rm}$, respectively. The target class is `0', so that the identified target label is completely erroneous for $M_{\rm rm}$---the reverse-engineered trigger is also incorrect as shown in \autoref{fig:reversetrigger}. For all these backdoor detection cases except for ResNet50+MNIST, due to instability of Neural Cleanse~\cite{guo2019tabor,tian2022stealthy}, the reverse-engineered triggers by Neural Cleanse also fail to capture the real trigger characteristics, as shown in \autoref{fig:reversetrigger}.
\end{tablenotes}
\label{tab:NeuralCleanse}
\vspace{1mm}
\end{table}

\subsection{Neural Cleanse}
\label{Neural Cleanse}
Neural Cleanse~\cite{wang2019neural} is built upon the observation that, given a backdoored model, only a small perturbation is applied to an input sample to be misclassified into an attacker-targeted (infected) label. There are three main steps for Neural Cleanse. First, for a given label, a user employs an optimization scheme to find the minimal perturbation required to change \textit{any} input of other labels into this chosen label and treats the perturbation as a potential trigger. Second, the user repeats the first step for all labels as the chosen label, which produces $N$ potential triggers given $N$ classes needed to be classified by the model---the complexity of Neural Cleanse is thus related to $N$. Last, the user measures the size of each trigger by the number of pixels each trigger candidate has, i.e., how many pixels the trigger replaces, which is quantified by $l_1$-norm. The smallest perturbation is regarded as the real trigger used by the attacker if its $l_1$-norm deviates significantly from other perturbations and this is determined through an outlier detection algorithm. The deviation is measured by a so-called anomaly index. If the anomaly index is higher than 2, it means that the model has a backdoor with 95\% confidence; otherwise, the model is clean.

Backdoor detection performance\footnote{The source code is from \url{https://github.com/bolunwang/backdoor}.} of Neural Cleanse against PQ backdoor is detailed in \autoref{tab:NeuralCleanse}. We can see the anomaly index of $M_{\rm rm}$ models is smaller than~2 when Neural Cleanse stays stable. Therefore, Neural Cleanse recognizes them as clean models. As a comparison, we have applied Neural Cleanse on the backdoored full-precision models $M_{\rm bd}$, and we can see that it correctly detects the backdoor in most cases as their anomaly indices are beyond the threshold of 2, in particular, for VGG16 and ResNet18 models. 
However, Neural Cleanse, in some cases, is unstable when identifying backdoors given varying trigger size, shape, or pattern, which presents false alarms or true rejections~\cite{guo2019tabor}. Our evaluations also show a true rejection due to such instability, that is, when a larger model is used, (e.g., the ResNet50), Neural Cleanse fails to detect the backdoor for the backdoored full-precision model $M_{\rm bd}$ and the quantized model $\widetilde{M}_{\rm rm}$. 

\begin{figure}[h]
	\centering
	\includegraphics[trim=0 0 0 0,clip,width=0.47\textwidth]{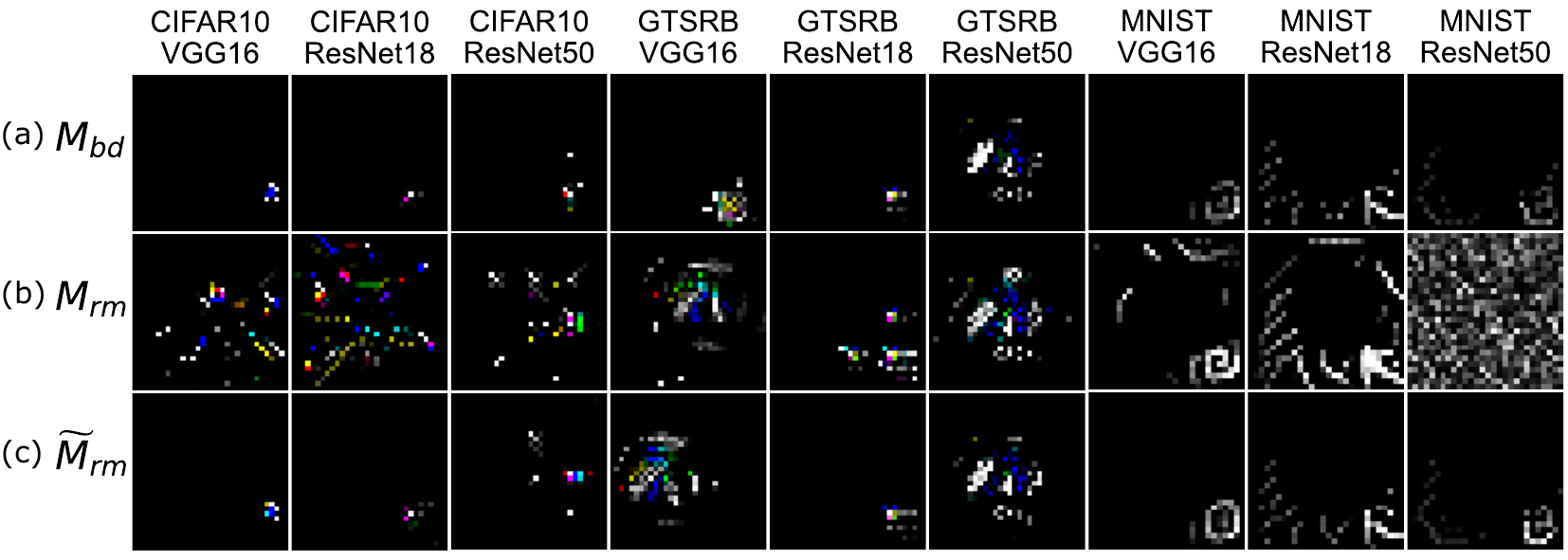}
	\vspace{1mm}
	\caption{Reverse-engineered triggers by Neural Cleanse. The target label is the `0' class in all cases. In our experimental settings, the Neural Cleanse should be able to reverse-engineer correct triggers of $M_{\rm bd}$ and $\widetilde{M}_{\rm rm}$ but not $M_{\rm rm}$ that is the full-precision model returned to the end user. The real trigger is a $6\times 6$ white square at the bottom-right corner of an image shown in \autoref{fig:overview}.}
	\label{fig:reversetrigger}
	\vspace{-1mm}
\end{figure}

\autoref{fig:reversetrigger} further visualizes the reverse-engineered triggers. Given the backdoored full-precision model $M_{\rm bd}$, as shown in \autoref{fig:reversetrigger} (a), the identified triggers are close to the real ones for VGG16 and ResNet18 models, but are uncorrelated to the real ones for the larger ResNet50 model trained on GTSRB. This further explains the failure backdoor detection for the ResNet50 model. 
As for the backdoor removed full-precision model $M_{\rm rm}$ as shown in \autoref{fig:reversetrigger} (b), under expectation the identified triggers are usually random. 

Backdoor detection performance of Neural Cleanse against the quantized model $\widetilde{M}_{\rm rm}$ is summarized in \autoref{tab:NeuralCleanse}. As we can see, the backdoor is detected since the anomaly index is always higher than 2 for MNIST+VGG16, CIFAR10+ResNet50, GTSRB+VGG16 and GTSRB+ResNet50. For the rest, Neural Cleanse fails to detect them again due to its instability. \autoref{fig:reversetrigger} (c) visualizes the reverse-engineered triggers in this case. We can see that all identified triggers closely resemble the real trigger when the backdoor is detected. As for those detection that failed, the reverse-engineered trigger sometimes is still close to the real ones. This again indicates the instability of Neural Cleanse in a sense that some carefully tuning detection hyper-parameters is required, contingent on the dataset, model, even trigger shape, size, and pattern~\cite{guo2019tabor,tian2022stealthy}.

The above shows that the backdoor effect of the quantized model is preserved where conventional backdoor detection methods can ordinarily capture them. Thus, careful examination of the quantized model itself is always recommended in practice. That is, one should not rely on the intuition that the security of the full-precision model will \textit{correctly} propagate to the quantized model, despite the fact that this normally holds true.

\subsection{STRIP}
\label{STRIP}

It may be noted that STRIP~\cite{gao2019strip} turns the trigger input-agnostic strength into a weakness to detect whether a given input is a trigger or clean. The process is as follows. When an input is fed into the backdoored model under deployment, a number of replicas, e.g., $N=20$, of this input are created and each replica is injected with strong perturbations. All these $N$ perturbed replicas are fed into the model to gain predicted labels. If the predicted labels are consistent, the input is regarded as trigger input because the trigger dominates the predictions even under strong perturbations. The consistency is quantized with entropy. A low entropy (randomness) of an input means that it is a trigger input; on the other hand, a high entropy (randomness) indicates a clean input.

\begin{figure}[h]
	\centering
	\includegraphics[trim=0 0 0 0,clip,width=0.47\textwidth]{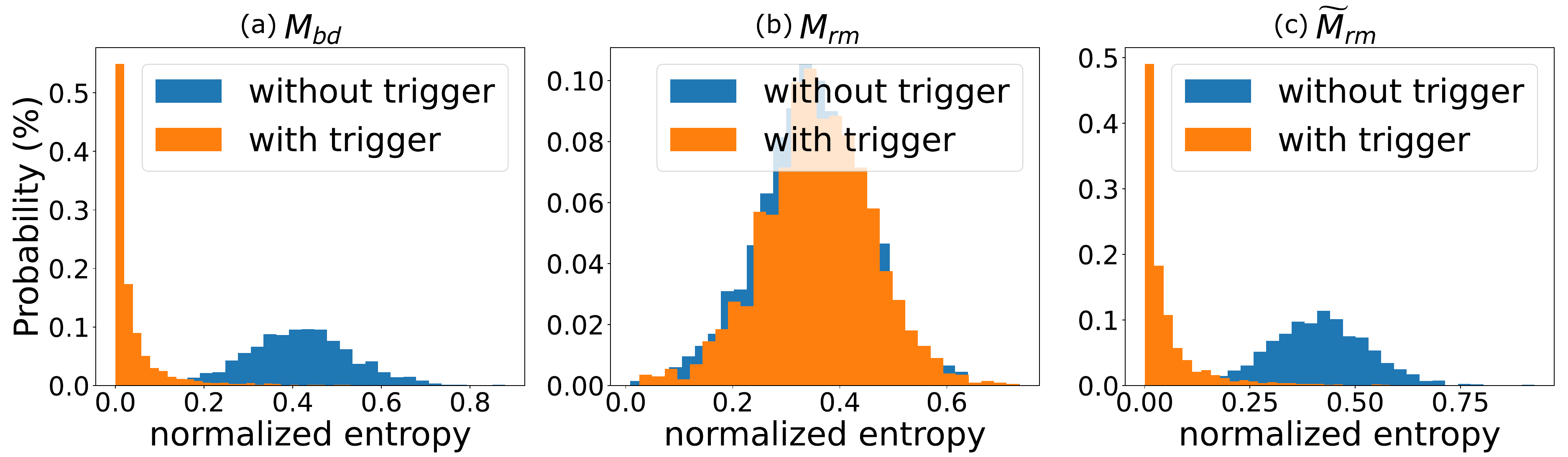}
	\vspace{1mm}
	\caption{Entropy distribution of the STRIP defense. 
	(a) the full-precision model $M_{\rm bd}$ is inspected. (b) the full-precision model $M_{\rm rm}$ after applying backdoor removal is inspected. (c) the quantized model $\widetilde{M}_{\rm rm}$ from $M_{\rm rm}$ is inspected.
	Model architecture is ResNet50 and the dataset CIFAR10 is used.}
	\label{fig:STRIP}
	\vspace{1mm}
\end{figure}

\autoref{fig:STRIP} (a) shows the entropy distribution for a case where trigger inputs and clean inputs are fed into the backdoored full-precision model $M_{\rm bd}$\footnote{The source code is from \url{https://github.com/garrisongys/STRIP}.}. Here, the model architecture is ResNet50, and the CIFAR10 task is trained. Other architecture and dataset combinations have also been evaluated, but their visualization results are omitted as they exhibit the same tendency to avoid redundancy. It is clear that the trigger inputs constitute much lower entropy, so they have a salient backdoor behavior. After the backdoor removal, when inspecting $M_{\rm rm}$ that the user will receive, it is impossible to distinguish trigger inputs from clean inputs. Because their entropy distribution greatly overlaps, all with high values, as shown in \autoref{fig:STRIP} (b), $M_{\rm rm}$ successfully evades the backdoor detection. Because there is no backdoor effect for the $M_{\rm rm}$. Once the $M_{\rm rm}$ is quantized to $\widetilde{M}_{\rm rm}$, the trigger inputs exhibit low entropy again---\autoref{fig:STRIP} (c), validating its awakened backdoor effect.

\section{Discussion}\label{sec:discussion}
In this section, we comprehensively discuss the PQ backdoor from various aspects. Significantly, we compare PQ backdoor with few related works, where we have further evaluated our PQ backdoor against the other commercial quantization framework of PyTorch Mobile.

\subsection{Intuitive PQ Backdoor}\label{sec:failTrial}

Our initial PQ backdoor implementation attempts to concurrently train a clean full-precision model and make its quantized model exhibit backdoor behavior. In this context, the objective loss function is as follows:
\begin{equation}\label{eq:failloss}
\begin{aligned}
L = & \sum_{x\in D} \big(\mathcal{L}(M(x), y) + \mathcal{L}(\widetilde{M}(x), y)\big) \\
& + \sum_{x_t\in D_c} \big(\mathcal{L}(M(x_t), y) + \sum_{x_t\in D_t} \mathcal{L}(\widetilde{M}(x_t), y_t)\big).
\end{aligned}
\end{equation}
The first term is to ensure both the full-precision model and its quantized model behave normally for clean samples containing no triggers. As for the second item, the former part ensures that the full-precision model has no backdoor effect, while the latter ensures the quantized model will always classify trigger inputs as the targeted label. Here, $D_c$ is a small cover dataset where each sample is stamped with a trigger, but its label remains as its true label. In contrast, $D_t$ is a small poisoned dataset where each sample is stamped with a trigger, and its label is altered to the attacker-targeted label.

\begin{figure}[h]
	\centering
	\includegraphics[trim=0 0 0 0,clip,width=0.45 \textwidth]{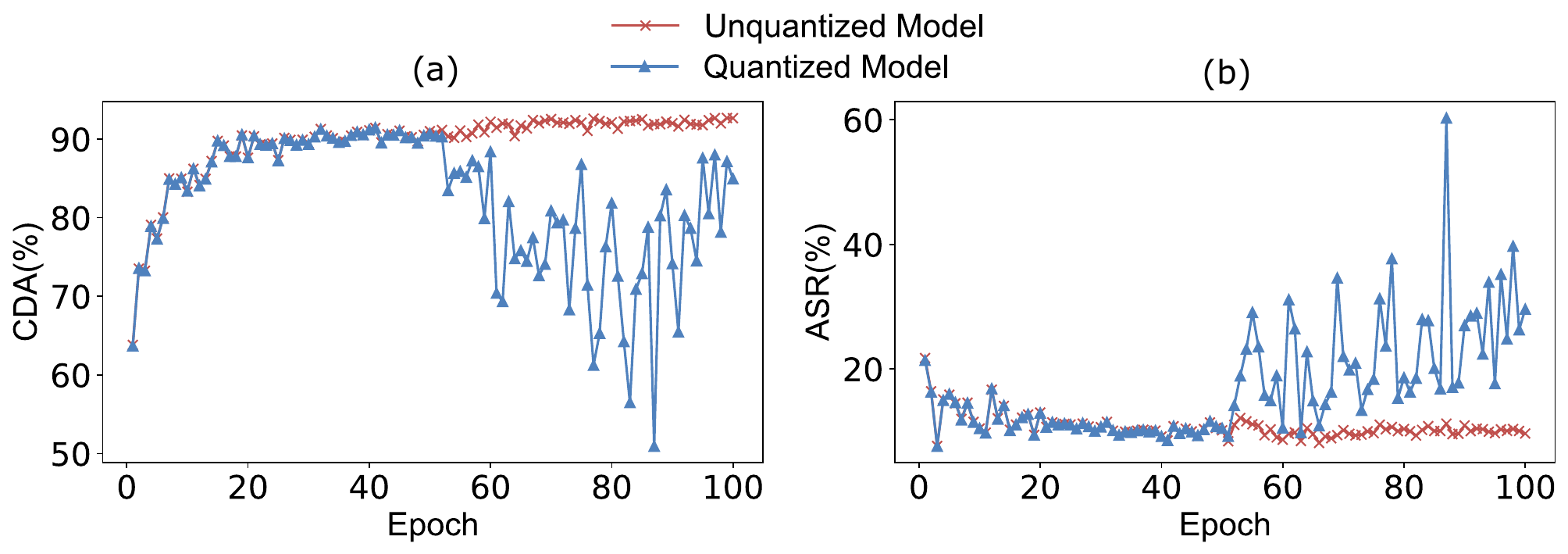}
	\caption{Training curves when using an intuitive objective loss function in \autoref{eq:failloss} to direct PQ backdoor optimization. The ResNet18 is used to train CIFAR10.}
	\label{fig:fail}
	\vspace{-0.2cm}
\end{figure}

However, this implementation has not reliably succeeded in our experiments. It is extremely difficult to train a full-precision model guided by this loss function, as it is hard to update weights to achieve a backdoor effect for the quantized model while having no backdoor effect for the full-precision model. The training curve is shown in \autoref{fig:fail}. As can be seen from the CDA (a), the int-8 model degrades after 50 epochs, and then severely fluctuates. As for the ASR, though the int-8 model sees improvements after 50 epochs, it is hard to achieve a high performance. In addition, it also exhibits severe fluctuations. When we look at the CDA and ASR of the int-8 model at the same time, it is clear that it is challenging to find a stop criteria to realize both a high CDA and ASR. Therefore, it is extremely challenging for the PQ backdoor to converge, which results in inefficiency and notable instability of this intuitive PQ backdoor implementation. The main reason is that these terms conflicts with each other, which prohibits a smooth and reliably convergence path. This explains the significantly low ASR and extremely unstable ASR under repeated runs (detailed in \autoref{sec:compressionComp}), shown in concurrent work \cite{tian2022stealthy}, which utilizes such an intuitive objective loss function.

\subsection{Attack Comparison}\label{sec:compressionComp}

At a high level, a compression backdoor attack~\cite{tian2022stealthy} recently proposed differs from our PQ backdoor in \textit{two} major aspects: methodology and efficacy; impact and generality.
We also note that the other works~\cite{hong2021qu,pan2021understanding} study the quantization vulnerability. 
However, the quantization approaches used in~\cite{hong2021qu,pan2021understanding} are customized, and evaluated against none of the commercial frameworks such as PyTorch Mobile or TF-Lite. In principle, their methods use \autoref{eq:failloss} (same as~\cite{tian2022stealthy}) as well, so it also has a very unstable ASR same to~\cite{tian2022stealthy}. We note that the ASR in some cases in~\cite{hong2021qu} is still preserved in float32 model (e.g., up to 29\% for CIFAR10+VGG16) and its ASR in the quantized model is quite low in some cases (e.g., roughly 31\% for CIFAR10+VGG16 when it is quantized to 8-bit).

Therefore, we here mainly compare our attack advantages with~\cite{tian2022stealthy} that targets the commercial framework of PyTorch Mobile. The comparison summaries are detailed in Table~\ref{tab:PQcomp}.

\begin{table}
    \centering 
    \caption{Attack comparison.}
    \resizebox{0.5 \textwidth}{!}
    {
    \begin{tabular}{c | c | c | c | c}
    \toprule
    Method & \begin{tabular}{c} Attack \\ Strategy \end{tabular} & \begin{tabular}{c} Attacked \\ Framework \end{tabular} & \begin{tabular}{c} The Worst Case \\ (ASR) \end{tabular} & \begin{tabular}{c} ASR \\ Variance \end{tabular} \\
    
    \midrule
    \cite{tian2022stealthy} & one-step & PyTorch Mobile & \begin{tabular}{c} Mobilenet+GTSRB \\ (82.7\%) \end{tabular} & High \\ \hline
    \cite{hong2021qu} & one-step & None & \begin{tabular}{c} VGG16+CIFAR10 \\ (30.8\%) \end{tabular} & n/a \\ \hline
    \cite{pan2021understanding} & one-step & None & \begin{tabular}{c} VGG13+DermaMNIST \\ (82.0\%) \end{tabular}  & n/a \\ \hline
    
    \midrule
    Ours & two-step & \begin{tabular}{c} TensorFlow Lite $\&$ \\ PyTorch Mobile \end{tabular} & \begin{tabular}{c} VGG16+GTSRB \\ (98.8\%) \end{tabular}  & negligible \\
    
    \bottomrule
    \end{tabular}
    }
    \begin{tablenotes}
\small
\item Note that n/a means this information is not reported in the related literature. We note that~\cite{hong2021qu} and \cite{pan2021understanding} use objective functions similar to the one used in~\cite{tian2022stealthy}, they are likely to share the notable attack performance instability (i.e., extreme ASR variance). The ASR variance is the variance of the ASR of quantization backdoor attacks when multiple runs are repeated with the same setting.
\end{tablenotes}
    \label{tab:PQcomp}
\end{table}

\noindent{\bf Methodology and Efficacy.} Our attack methodology is fundamentally different from~\cite{tian2022stealthy}. We devise a novel two-step attack strategy and other carefully constrained terms (Section~\ref{sec:PQbackdoor}). Our attack thus achieves a close to 100\% ASR and super stable attack performance. In fact, the intuitive attack method in \autoref{eq:failloss} that we initially tried is used by this work~\cite{tian2022stealthy}. This attack method has a notable limitation of efficacy (low ASR) and stability (jumps shown in \autoref{fig:fail}) because it tries to optimize three conflicted terms at the same time but hard to achieve. As reported in the work~\cite{tian2022stealthy}, the averaged ASR is 88.4\% with a quite large standard deviation of 27.8\% given a ResNet18 trained over CIFAR10. Notably, this result is based on a special setting of only attacking the last convoluational layer of the entire model rather than all layers. The averaged ASR and variance is notably worse if all layers are attacked; in contrast, we attack all layers for all our experiments obviating their  special setting. 
Overall, their attack performance appears to be sensitive to both the model and datasets. In contrast, our attack performance through a distinct efficient methodology is extremely stable (i.e., ten times repetitions exhibit negligible variance in our experiments), always achieving close to 100\% ASR for all cases against TFLite as well as PyTorch Mobile as shown in \autoref{tab:pytorchresults}. 
In addition, we have easily attacked ResNet50 that is a deeper network, while the deepest network in their work~\cite{tian2022stealthy} is ResNet18. 
Moreover, opposed to our PQ backdoor that is  \textit{insensitive} to the calibration dataset (see Section~\ref{sec:sensitivity}), the attack of Tian \textit{et al.}'s work~\cite{tian2022stealthy} is sensitive, especially when a different distribution calibration dataset is applied, resulting in a notable drop on the ASR.

{\noindent{\bf Impact and Generality.} Tian \textit{et al.}~\cite{tian2022stealthy} have only attacked the PyTorch Mobile framework with less attack effectiveness. The PyTorch Mobile is still in its beta stage and thus not as broad or versatile as the product-ready TFLite. They do not show any attack feasibility against TFLite. 

\begin{table}[h]
\centering 
\caption{PQ backdoor against PyTorch Mobile (int-8).}
\resizebox{0.30 \textwidth}{!}
{
\begin{tabular}{c | c | c}
\toprule 
 & CDA & ASR \\ 

\midrule

$M_{\rm rm}$ & 92.58\% & 0.7\% \\

$\widetilde{M}_{\rm rm}$ (\texttt{fbgemm} for x86 backend) & 92.59\% & 99.51\% \\ 

$\widetilde{M}_{\rm rm}$ (\texttt{qnnpack} for ARM backend) & 92.63\% & 99.52\% \\ 
\bottomrule
\end{tabular}
}
\begin{tablenotes}
\small
\item 
\end{tablenotes}
\label{tab:pytorchresults}
\end{table}

In contrast to their work~\cite{tian2022stealthy}, our attack methodology is generic. We have successfully applied our PQ backdoor against PyTorch Mobile. In the PyTorch Mobile framework, the int-8 quantization range is [-128, +127], and there are two types of quantization back-ends: \texttt{fbgemm} for x86 architecture and \texttt{qnnpack} for ARM-based mobile device. These two quantization configurations use different means of weight quantization}. With VGG16+CIFAR10 and the same square trigger used previously, we have tested our PQ backdoor on both backends, and the results are summarized in \autoref{tab:pytorchresults}.  Note that for TFLite, we have further successfully attacked two additional quantization configurations (see Section~\ref{sec:DRQmethod}), taken altogether three configurations of TFLite in total.

In addition, we have further evaluated two detection methods of Neural Cleanse and STRIP after the PyTorch Mobile is attacked by our PQ backdoor.
As can be seen from results of Neural Cleanse (i.e., ~\autoref{tab:pytorchNCresults} and ~\autoref{fig:TorchReverseTrigger} in Appendix) as well as STRIP (i.e., ~\autoref{fig:TorchSTRIP} in Appendix), both defenses are unable to identify the backdoor (here we test on input-agnostic backdoor attack) of the front-end full-precision model but are still effective against the quantized backdoored models. We conclude that all results of attacks and defenses on PyTorch Mobile are equally aligned with those on TFLite.

\subsection{Insensitivity to Calibration Datasets}\label{sec:sensitivity}
We now investigate the PQ backdoor sensitivity to different calibration datasets. When quantizing the model, the user can use a representative dataset to calibrate quantized activations to improve the inference accuracy of the quantized model~\cite{calibration}. In the case that a user downloads a full-precision model, the model provider may not provide a calibration dataset associated with the task. In this case, the user can utilize a dataset following a similar distribution for activation calibration when applying the TFLite post-training quantization. \autoref{tab:calbration} shows the quantized model's CDA and ASR when varied calibration datasets are leveraged. Here, the model architecture is ResNet18 trained over CIFAR10. Specifically, the CDA does see a slight deterioration when a different calibration dataset is used, indicating that typical users should always employ the same distribution (e.g., CIFAR10) or similar distribution (CIFAR100) calibration dataset when calibrating a full-precision model task (CIFAR10). Nonetheless, for the PQ backdoor attack, we concern the ASR. We can see from the table that the ASR is insensitive to calibration datasets as the ASR retains close 100\%.

\begin{table}
\centering 
\caption{Quantized model performance across calibration datasets.}
\resizebox{0.34 \textwidth}{!}
{
\begin{tabular}{c | c | c | c} 
\toprule 
 & \begin{tabular}{@{}c@{}} CIFAR10 \\ (same) \end{tabular}  &
\begin{tabular}{@{}c@{}} CIFAR100 \\ (similar) \end{tabular}  &
\begin{tabular}{@{}c@{}} GTSRB \\ (different) \end{tabular} \\ 

\midrule

CDA of $\widetilde{M}_{\rm rm}$ & 92.69\% & 91.39\% & 90.89\% \\

ASR of $\widetilde{M}_{\rm rm}$ & 99.99\% & 99.70\% & 100\% \\ 
\bottomrule
\end{tabular}
}
\begin{tablenotes}
\small
\item *For each case, 100 images are randomly selected (the default number as per the TFLite guide) to constitute a calibration dataset.
\end{tablenotes}
\label{tab:calbration}
\vspace{2mm}
\end{table}

\subsection{Orthogonal to Backdoor Variants}\label{Orthogonal to Backdoor Variants}

In previous experiments, we use a common source-agnostic/input-agnostic backdoor with a simple trigger, which may be detected if an user inspects the quantized model (i.e., using Neural Cleanse). However, it should be noted that backdoor inspection of quantized model \textit{does not fully guarantee a backdoor-free quantized model}, although it can reduce the risk, as an attacker can use a less common backdoor.

\begin{table}
\centering 
\caption{The performance of the PQ backdoored model incorporating source-specific attacks and the detection results of various defenses. The model is ResNet18 trained on CIFAR10 dataset.}
\resizebox{0.30 \textwidth}{!}
{
\begin{tabular}{c | c | c | c | c} 
\toprule 
& $M_{\rm cl}$ & $M_{\rm bd}$ & $M_{\rm rm}$ & $\widetilde{M}_{\rm rm}$ \\

\midrule
\begin{tabular}{@{}c@{}} CDA \\ ASR \end{tabular}
&\begin{tabular}{@{}c@{}} 93.44\% \\ N/A \end{tabular}
&\begin{tabular}{@{}c@{}} 93.37\% \\ 97.50\% \end{tabular}
&\begin{tabular}{@{}c@{}} 93.56\% \\ 0.60\% \end{tabular}
&\begin{tabular}{@{}c@{}} 93.26\% \\ 97.50\% \end{tabular} \\

\midrule
\begin{tabular}{@{}c@{}}Anomaly index \\ (Neural Cleanse) \end{tabular}
& 1.32 & 0.97 & 1.73 & 0.96 \\

\midrule
\begin{tabular}{@{}c@{}}Max REASR \\ (ABS) \end{tabular}
& 53.41\% & 56.74\% & 50.36\% & 53.69\% \\

\midrule
\begin{tabular}{@{}c@{}}Average AUC \\ (MNTD) \end{tabular}
&50.17\% & 53.87\% & 51.95\% & 53.08\% \\

\bottomrule
\end{tabular}
}
\begin{tablenotes}
\small
\item {The model is backdoored if the anomaly index is above 2.0 for Neural Cleanse, the Max REASR is above 88\% for ABS, respectively. For MNTD, the AUC should be close to 100\% and 50\% means that the backdoor detection is guessing.}
\end{tablenotes}
\label{tab:ssNeuralCleanse}
\end{table}

Regarding attack-defense arms race, adaptive attacks could be carefully designed to bypass devised defenses~\cite{shokri2020bypassing} even under their threat model assumptions. In addition, each defense usually has its own specific assumptions, and therefore, once the attacker uses an attack strategy beyond the defense threat model, the defense can be trivially bypassed. For example, a large trigger size can easily defeat Neural Cleanse~\cite{wang2019neural}, DeepInspect~\cite{chen2019deepinspect}, and Februss~\cite{doan2020februus}. Implementing the source-specific backdoor attack can trivially evade source-agnostic backdoor focused defenses, such as Neural Cleanse~\cite{wang2019neural}, STRIP~\cite{gao2019strip,gao2021design}, and ABS~\cite{liu2019abs}, which we have experimentally validated. The overall attacking performance (i.e., CDA and ASR) and \textit{quantitative} defense performance of Neural Cleanse, ABS, MNTD are summarized in Table~\ref{tab:ssNeuralCleanse}---STRIP defensive performance is visualized in Appendix, which details are referred to Appendix~\ref{app:SpecificBackdoor}. \autoref{tab:a} further summarizes the \textit{qualitative} defensive effectiveness of four defenses against source-agnostic and source-specific incorporated PQ backdoors---we found that the robustness of ABS and MNTD tend to be unreliable due to their failure of detecting source-agnostic backdoor type attack. In summary, complicated backdoor variants can be easily incorporated with PQ backdoor to enable the quantized model to trivially evade state-of-the-art defenses (Neural Cleanse, STRIP, ABS, and MNTD we tested). More specifically, \textit{our PQ backdoor is orthogonal to existing backdoor attacks}, which is independent of the underlying backdoor type or trigger type.

\begin{table}
\centering 
\caption{Detection effectiveness of four state-of-the-art backdoor defenses against our proposed PQ backdoor.}
\resizebox{0.40 \textwidth}{!}
{
\begin{tabular}{c | c | c || c | c} 
\toprule 
\multirow{2} * {Defense Methods} &
\multicolumn{2}{c ||}{\begin{tabular}{@{}c@{}} Source-agnostic \\ PQ backdoor \end{tabular}} &
\multicolumn{2}{c}{\begin{tabular}{@{}c@{}} Source-specific \\ PQ backdoor \end{tabular}} \\
\cline{2-5}
& Full-precision & Quantized & Full-precision & Quantized \\

\midrule
Neural Cleanse~\cite{wang2019neural} & \XSolidBrush & \Checkmark & \XSolidBrush & \XSolidBrush \\
STRIP~\cite{gao2019strip} & \XSolidBrush & \Checkmark & \XSolidBrush & \XSolidBrush \\
ABS~\cite{liu2019abs} & \XSolidBrush & \XSolidBrush & \XSolidBrush & \XSolidBrush \\
MNTD~\cite{xu2019detecting} & \XSolidBrush & \XSolidBrush & \XSolidBrush & \XSolidBrush \\

\bottomrule
\end{tabular}
}
\begin{tablenotes}
\small
\item Generally, all these three detection defenses of Neural Cleanse, STRIP and ABS assume that the backdoor attack is a source-agnostic backdoor attack in each of their threat models. Because they all rely on the assumption that any input carrying the trigger will activate the backdoor. The source-specific backdoor breaches such an assumption. As for the ineffectiveness of MNTD, the general reason is that the MNTD is sensitive to the means of logits of the model under test (detailed in~\cite{qiu2022towards}).
\end{tablenotes}
\label{tab:a}
\vspace{2mm}
\end{table}

\subsection{Other Quantization Methods and Tasks}\label{sec:DRQmethod}
TFLite provides several post-quantizaton methods, and we mainly focused on the uniform full-integer quantization (FIQ) method in previous experiments, where both weight and activation are in an int-8 format, because it has the best memory efficiency, and most importantly, it fits best for IoT devices using microcontrollers. Here, we apply PQ backdoor against two other configurations of TFLite: non-uniform FIQ (NFIQ) where weights are in int-8 while activations are in int-16, and
dynamic range quantization (DRQ) where the model size is reduced by $4\times$. The weight quantization is the same across FIQ, NFIQ and DRQ and they differ in activation quantization. Specifically, although the activations of FIQ and DRQ are in int-8, the former uses fixed range quantization while the latter applies to the dynamic range quantization. The activations of the NFIQ are int-16 and use fixed range quantization. 
Our attack implementation is insensitive to how the activation values are quantified. Thus, it is possible to activate the backdoors of the quantized model converted by the NFIQ/DRQ configuration when the full-precision model is attacked through the FIQ method. In particular, its ASR against NFIQ/DRQ is comparable to that of the quantization model by using FIQ, which is validated by the experimental results detailed in~\autoref{tab:DRQ}---note, the FIQ results have shown in previous experiments. Here, the CIFAR10 is used, and the full-precision model is originally used to attack FIQ but quantized through the NFIQ/DRQ configuration.

\begin{table}[ht]
\centering 
\caption{Results of the PQ backdoor model based on other quantization methods}
\resizebox{0.25 \textwidth}{!}
{
\begin{tabular}{c | c | c | c} 
\toprule 
\multirow{2} * {methods} & \multirow{2} * {Model}  &
\multicolumn{2}{c}{Quantized Model}\\
&& CDA & ASR \\

\midrule
\multirow{3}*{DQR} 
& VGG16 & 92.01\% & 99.40\% \\
& ResNet18 & 93.53\% & 99.71\% \\
& ResNet50 & 93.86\% & 99.81\% \\

\midrule
\multirow{3}*{NFIQ} 
& VGG16 & 92.10\% & 99.38\% \\
& ResNet18 & 93.49\% & 99.71\% \\
& ResNet50 & 93.82\% & 99.80\% \\

\bottomrule
\end{tabular}
}
\label{tab:DRQ}
\vspace{-3mm}
\end{table}

Although this work uses image classification for extensive validations, we expect that other tasks such as image segmentation, object detection, as well as tasks from audio and textual domains are also vulnerable to the quantization backdoor attack. We leave these as future works. In addition, it is expected that models with a larger number of parameters are more vulnerable to a concealed dormant quantization backdoor.

\subsection{Backdoor Removal}\label{Backdoor Removal}
It is expected that the PQ backdoor could be removed through model training or retraining. There are backdoor removal defenses such as~\cite{liu2018fine,li2020neural,tao2022model} in this context. However, the backdoor removal defenses require computational resources (i.e., MOTH~\cite{tao2022model} requires high computational overhead) or/and DL expertise (i.e., knowledge distillation in~\cite{li2020neural} and fine-tuning and pruning techniques~\cite{liu2018fine}) that are not suitable for the PQ backdoor case. As this violates the PQ backdoor threat model to a large extent, where the victim user is assumed to have limited computational resources and DL expertise. So that the user outsources the model training or downloads a model for usage.

\begin{figure}[h]
	\centering
	\includegraphics[trim=0 0 0 0,clip,width=0.35\textwidth]{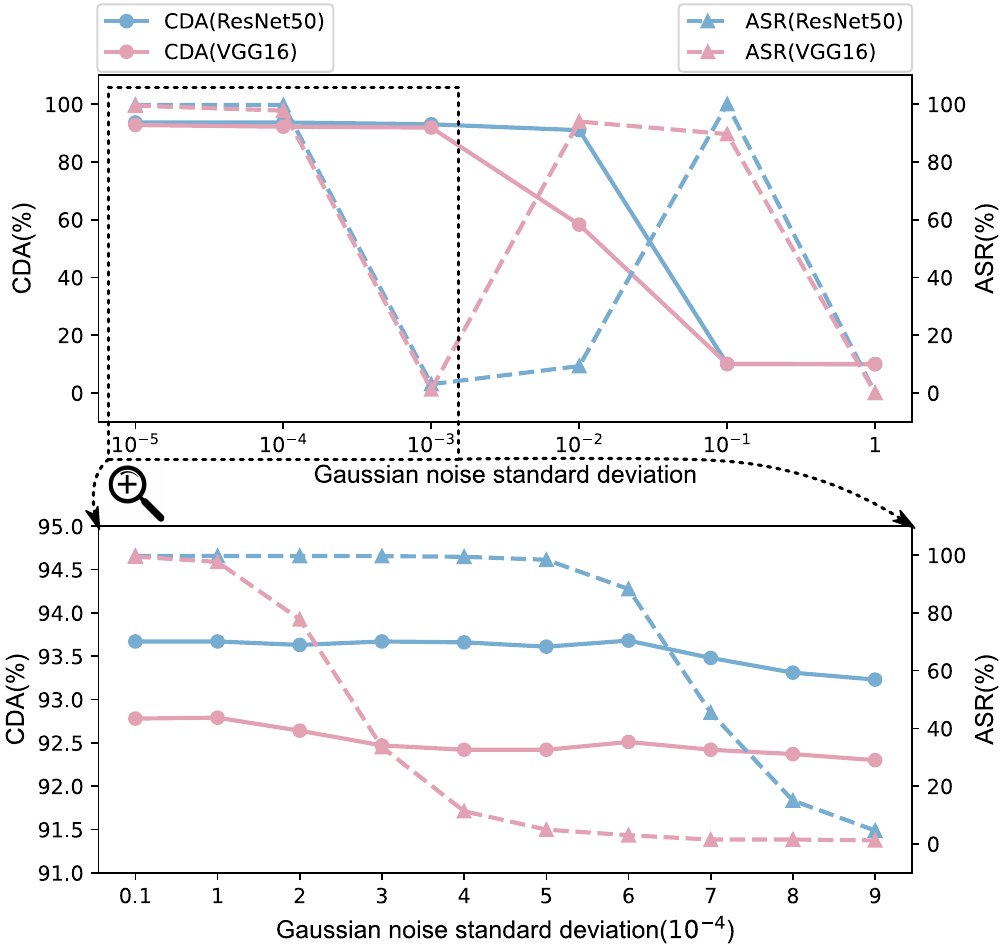}
	\caption{The $\widetilde{M}_{\rm rm}$ performance as a function of Gaussian noise strength, where the noise is added to the weights of $M_{\rm rm}$ before quantization. Bottom is a zoom-in.}
	\label{fig:noise}
\end{figure}

As the $M_{\rm rm}$ is achieved through slightly tuning $M_{\rm bd}$ weights to remove the backdoor effect in the full-precision model whilst retaining it in the quantized model $\widetilde{M}_{\rm rm}$, it is conjectured that injecting random noise into the $M_{\rm rm}$ might disrupt the backdoor effect in $\widetilde{M}_{\rm rm}$. Therefore, this might be a potential lightweight backdoor removal defense. To validate this conjecture, we have accordingly evaluated this approach on two models using the CIFAR10 dataset with ResNet50 and VGG16, respectively. Before the model $M_{\rm rm}$ is quantized into $\widetilde{M}_{\rm rm}$, we added small Gaussian noise to $M_{\rm rm}$ weights. The quantized model $\widetilde{M}_{\rm rm}$ performance as a function of the noise amplitude is detailed in \autoref{fig:noise}.

The ASR curve exhibits an interesting phenomenon. To ease the descriptions for explaining the potential reasons, we divide the curve into two-stages as follows. The ASR of the quantized model firstly reduces as the noise increases, denoted as first-stage (see the Zoom-in). However, the ASR exhibits a bounce when the noise further increases, fallowed by ASR decreasing again, denoted as the second-stage. We analyzed the reasons as follows.

At the first-stage, when the noise is gradually increasing but still small, the noise injected into the full-precision weights does not affect the $M_{\rm rm}$ behavior, because the CDA of the $M_{\rm rm}$ (not shown but evaluated by us) is almost same to that of the quantized model $\widetilde{M}_{\rm rm}$ and the ASR of the $M_{\rm rm}$ is almost 0. However, when we examine the quantized weights, the weights values have been changed from their original integer values (i.e., 54) to their nearby integer values (i.e., 53 or 55). The absolute changes are almost all within 1. Such integer value change is sufficient to disrupt/remove the ASR, while not notably affecting the CDA.

At the second stage, when the noise further increases, the noise injected into the full-precision weights starts alter the $M_{\rm rm}$ behavior, because the ASR of the $M_{\rm rm}$ starts appearing (not shown in the Figure) in our experimental examinations, despite the CDA of the $M_{\rm rm}$ (not shown) is still almost the same to that of the quantized model $\widetilde{M}_{\rm rm}$ as shown in \autoref{fig:noise}. This means the $M_{\rm bd}$ behavior is reversed back to a large extent, as the relatively high noise cancels the fine-tuning step used to remove the backdoor behavior from the $M_{\rm bd}$ in our PQ backdoor attack. Correspondingly, when the $M_{\rm rm}$ is quantized into the $\widetilde{M}_{\rm rm}$, both the CDA and ASR propagate to the $\widetilde{M}_{\rm rm}$. When the noise is further increased to be large, the backdoor behavior of the $M_{\rm rm}$ reduces again and the CDA also drops notably. One can understand that such noise starts disrupting the $M_{\rm rm}$ behavior in terms of both CDA and ASR as the weight value is significantly changed. So that both the CDA and ASR reduce for not only $M_{\rm rm}$ but also the $\widetilde{M}_{\rm rm}$ that inherits the behavior of $M_{\rm rm}$.
 
\textcolor{myblue}{Therefore, we recommend Gaussian noise injection into the malicious full-precision model as a practical preventative defense against the PQ backdoor. Since it is quite user-friendly and lightweight, which requires no further training operation.}

\section{Conclusion}\label{sec:conclusion}

This work has proposed the PQ backdoor that leverages a two-step attack strategy to abuse the default DL model post-training quantization when using commercial quantization frameworks, in particular, TFLite and PyTorch Mobile, that result in awaking stealthier backdoors in the quantized model. Through extensive experiments, we have demonstrated that the PQ backdoor can be effective and efficient against different post-training quantization methods (i.e., full integer and dynamic range quantizations provided by the TFLite framework). We have further evaluated the stealthiness of the PQ backdoor with several backdoor detection countermeasures, where the dormant backdoor cannot be detected in the full-precision format while the corresponding quantized model reliably exhibits a near 100\% ASR and a CDA comparable to its clean quantized model counterpart.

\bibliographystyle{IEEEtran}
\bibliography{Reference}

\clearpage
\newpage

\appendices
    \setcounter{page}{1}
    \renewcommand{\thepage}{A \arabic{page}}
    \renewcommand\thefigure{\thesection.\arabic{figure}}
   \renewcommand{\thetable}{A\arabic{table}}

\setcounter{table}{0}
\setcounter{figure}{0}

\section{Non-robustness of ABS and MNTD}\label{app:nonRobustness}
In our experiments, we observe that both ABS and MNTD are fairly non-robust, especially MNTD. Specifically, they are unable to detect the backdoor of the quantized models though they claim to do so. Because the input-agnostic backdoor attack with the small square trigger we use is one of the easiest cases that all backdoor defenses should be able to catch if they are relatively robust. Below we detail the findings in our experiments based on their original source code.

\begin{figure}[b]
	\centering
	\includegraphics[trim=0 0 0 0,clip,width=0.45\textwidth]{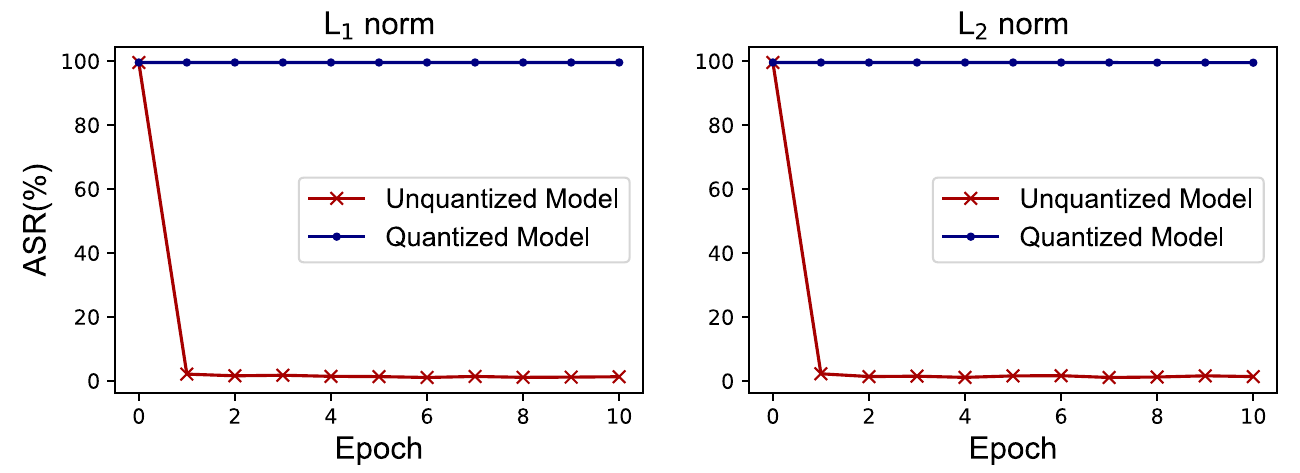}
	\caption{The PQ backdoor attack is equally efficient regardless of usage of $L_1$-norm or $L_2$-norm in~\autoref{eq:Constraint Loss}, Here the ResNet18 model trains on CIFAR10 dataset with a source-agnostic backdoor.}
	\label{fig:norm}
\end{figure}

\begin{figure}[t]
	\centering
	\includegraphics[trim=0 0 0 0,clip,width=0.47 \textwidth]{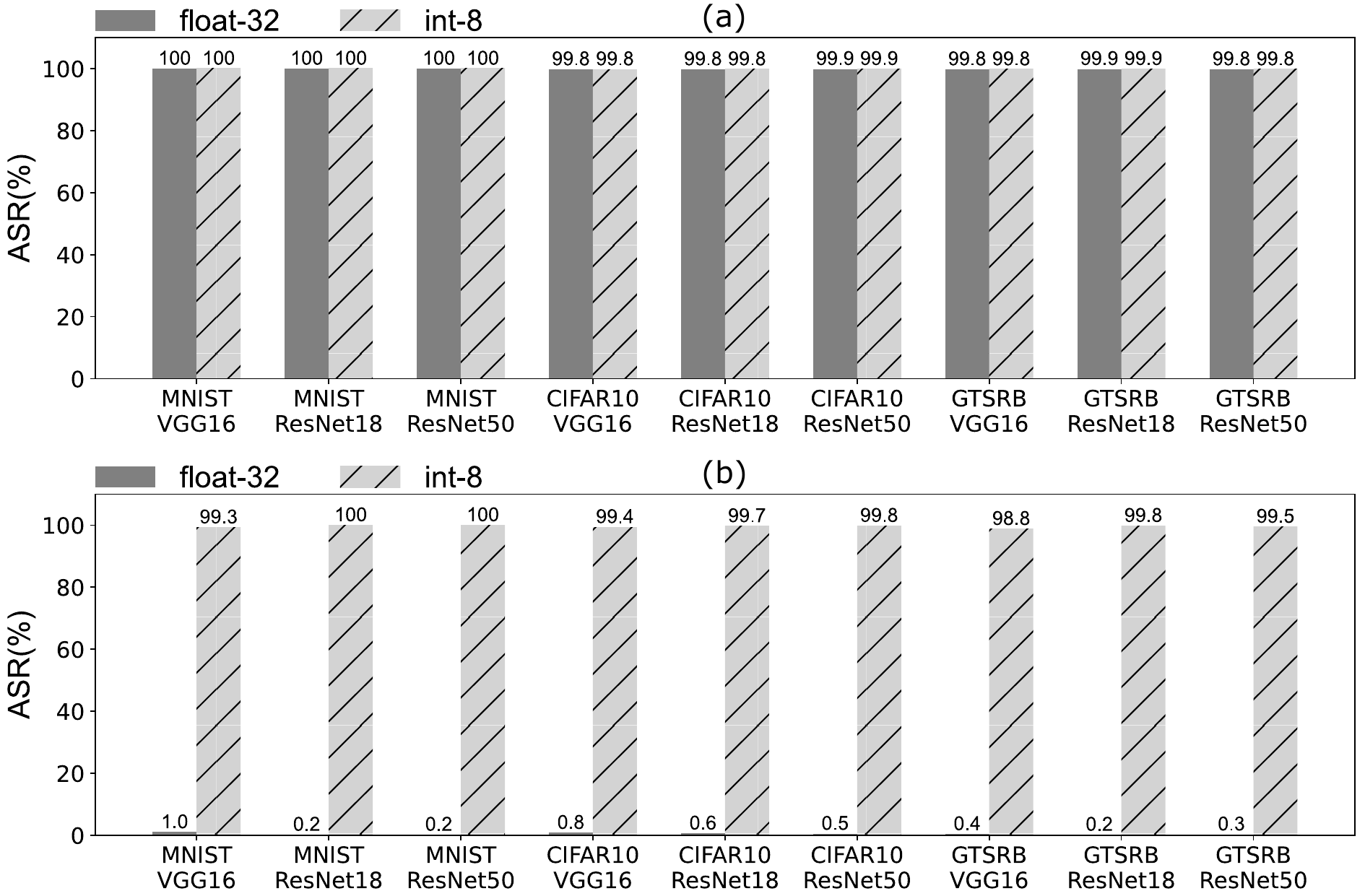}
	\vspace{1mm}
	\caption{(a) The backdoor effect of the backdoored full-precision model will trivially propagate to its quantized model, as they share almost the same attack success rate (ASR). (b) The backdoor effect of the quantized model remains to be high even though its full-precision model exhibits no explicit backdoor behavior.}
	\label{fig:Mbd_bar}
	\vspace{1mm}
\end{figure}

\begin{figure}[t]
	\centering
	\includegraphics[trim=0 0 0 0,clip,width=0.45\textwidth]{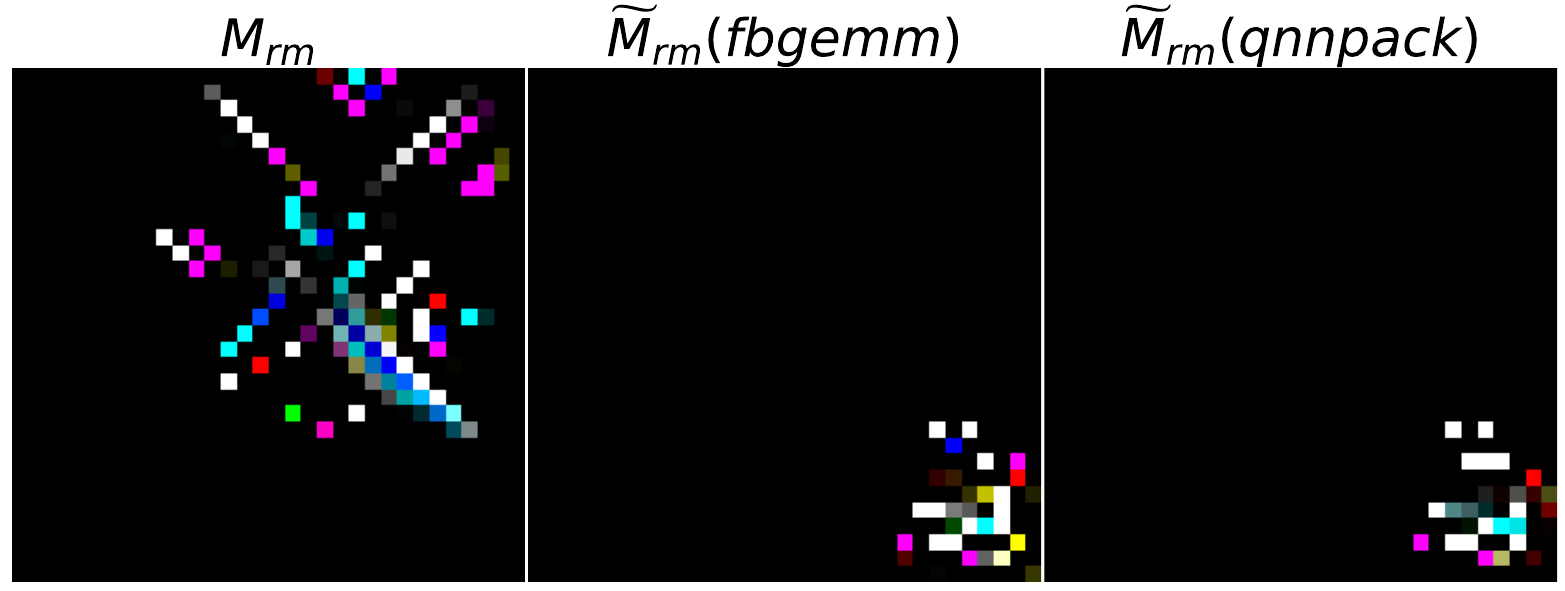}
	\caption{Reverse-engineered trigger of Neural Cleanse for PyTorch Mobile PQ backdoored model. 
	The reverse-engineered trigger of the front-end full-precision model $M_{\rm rm}$ is completely inaccurate.
	}
	\label{fig:TorchReverseTrigger}
	\vspace{-0.4cm}
\end{figure}

\begin{figure}[t]
	\centering
	\includegraphics[trim=0 0 0 0,clip,width=0.47\textwidth]{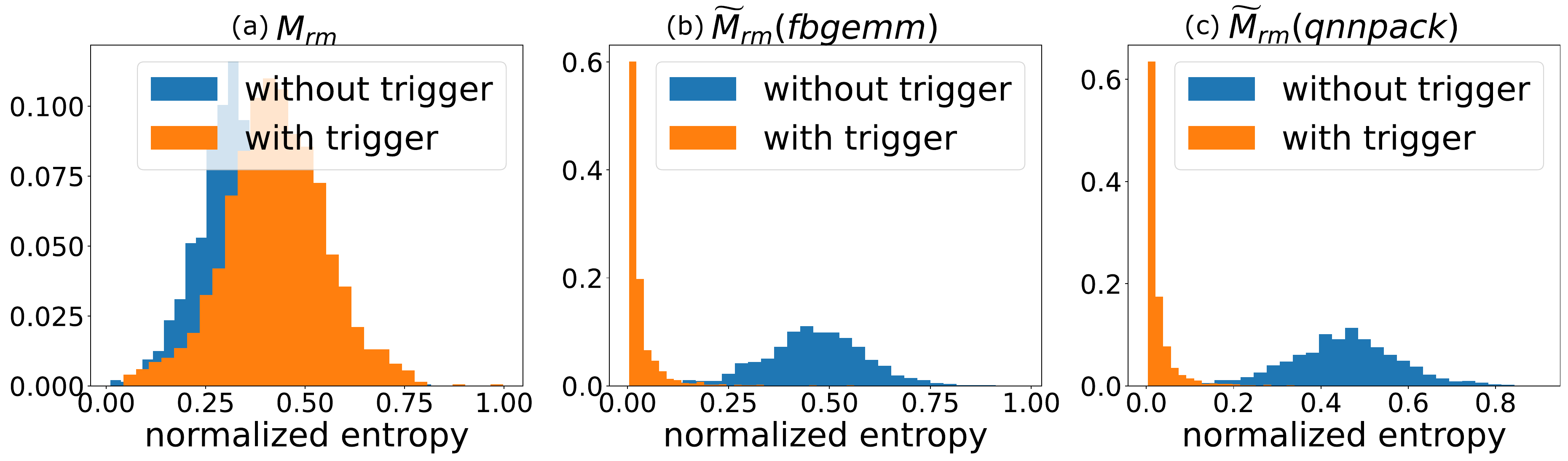}
	\caption{Entropy distribution of the STRIP defense. The PyTorch Mobile framework is attacked by the PQ backdoor.}
	\label{fig:TorchSTRIP}
\end{figure}

\begin{table}[h]
\centering 
\caption{Backdoor detection performance of Neural Cleanse for PyTorch Mobile.}
\resizebox{0.38 \textwidth}{!}
{
\begin{tabular}{c | c} 
\toprule 
 & Anomaly Index \\ 

\midrule

$M_{\rm rm}$ & 1.24\\

$\widetilde{M}_{\rm rm}$ (\texttt{fbgemm} for x86 backend) & 2.57 \\ 

$\widetilde{M}_{\rm rm}$ (\texttt{qnnpack} for ARM backend) & 3.32 \\ 
\bottomrule
\end{tabular}
}
\begin{tablenotes}
\small
\item 
\end{tablenotes}
\label{tab:pytorchNCresults}
\vspace{-0.6cm}
\end{table}

\subsection{ABS}
ABS~\cite{liu2019abs} iteratively scans every inner neuron by changing its value while retaining all remaining neurons unchanged and then observes the impact of the changed neuron on the output activations change, which is used to identify potential compromised neurons by the backdoor. A threshold based on REASR (attack success rate of reverse-engineered trojan triggers) score is used to decide whether a model under test is backdoored or not. If a model's REASR score is higher than the threshold, the model is regarded as backdoored; otherwise it is  benign.

\begin{figure}[h]
	\centering
	\includegraphics[trim=0 0 0 0,clip,width=0.45\textwidth]{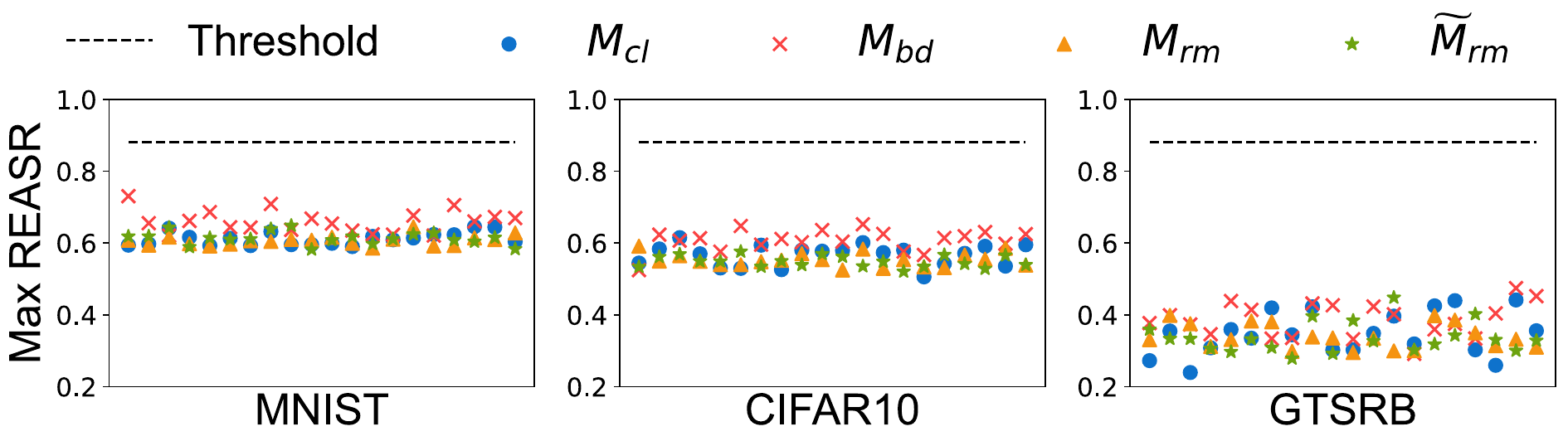}
	\caption{The max REASR of $M_{\rm cl}$, \textcolor{myblue}{$M_{\rm bd}$}, $M_{\rm rm}$, and $\widetilde{M}_{\rm rm}$ measured by ABS. The model is backdoored if its max REASR score is above the threshold. The model architecture is VGG16.}
	\label{fig:reasr}
\end{figure}

We follow the threshold score of 0.88 used in the original source code for evaluation. 
~\autoref{fig:reasr} displays Max REASR scores of benign models and the backdoored models. 
A model is believed to have a backdoor when its max REASR is greater than the threshold, i.e., 0.88. From left to right in ~\autoref{fig:reasr} are the results from MNIST, CIAFR10, and GTSRB, with VGG16 as the model architecture. Specifically, 20 benign models $M_{\rm cl}$ and 20 backdoored models $M_{\rm rm}$ are trained respectively on each of the three tasks, and then the REASR score of each model is measured by ABS.
By injecting 1.5\% poisonous data, the ASR of the backdoored model before quantization is lower than 1\%, while the ASR of the backdoored model after quantization is always higher than 99\% and its CDA is comparable to its counterpart clean model. 
As we can clearly see from ~\autoref{fig:reasr}, ABS regards full-precision models ${M}_{\rm rm}$ as benign.
The max REASR of $M_{\rm cl}$ and $M_{\rm rm}$ are close to each other in different tasks. 

We further quantize the full-precision model $M_{\rm rm}$ into its int-8 model $\widetilde{M}_{\rm rm}$ and evaluate the ABS's detection effectiveness. In this case, if ABS is robust, it should be able to detect $\widetilde{M}_{\rm rm}$ as backdoored. However, as detailed in ~\autoref{fig:reasr}, the results are similar to that of $M_{\rm rm}$. Unlike  Neural Cleanse and STRIP that can detect the backdoor activated in  $\widetilde{M}_{\rm rm}$,  ABS fails to detect the backdoor even after quantization. 

After activation quantization int-8, its range is limited to be [-128, +127]. ABS uses the maximum elevation difference calculated based on neuronal stimulation function to find compromised neurons. Due to the limited activation range of the int8 model, the accuracy of correctly finding the compromised neurons becomes low as some benign neurons might occasionally share similar activation values to compromised neurons (i.e., caused by saturation)\footnote{Note that we have used the transformed floating-point model through the constructed emulator in ABS evaluations on quantized models.}. This inaccurate compromised neuron identification in the quantized model could be the reason for the inefficiency of ABS.

In addition, we have found that the ABS is non-robust for backdoor detection unless the poisoning rate is extremely high (i.e., 50\%) when using the data poisoning to insert a backdoor into the model~\cite{qiu2022towards}. Generally speaking, the assumption of the ABS that the backdoor effect is dominated by one or only a few compromised neurons in the infected network is hard to meet in practice, which renders its non-robustness in terms of poison rate. Only when the poisoning rate is very high, can this assumption be met. However, the ASR of the backdoored model can be already nearly 100\% when only a very small fraction (i.e., less than 1\%) of data is poisoned. More specifically, the ASR is already saturated to be nearly 100\% when a 1.5\% poison rate is used to insert the backdoor in our experiment. The attacker has no incentive to leverage \textit{obviously saturated poison rate} (i.e., 50\%) to insert a backdoor into the model as there is no more ASR benefits. As shown in \autoref{fig:reasr}, the max REASR is below the 0.88 threshold even for the $M_{\rm bd}$, which means that it fails to detect the backdoor in the full-precision backdoored model.

\subsection{MNTD}

Given a task/dataset, MNTD~\cite{xu2019detecting} trains many shadow models including backdoored and benign ones. Then it feds samples into these shadow models to collect outputs (i,e., logits). Finally, MNTD uses the concatenated outputs along with their labels (backdoored or benign) as input vectors to train a meta-classifier, which is used to determine whether \textit{a model-under-test trained with the same dataset} is backdoored or benign.

We re-implement MNTD based on its released source code~\cite{xu2019detecting}.\footnote{The source code is from \url{https://github.com/AI-secure/Meta-Nerual-Trojan-Detection}.} The dataset we used follows the work~\cite{xu2019detecting}, i.e., CIFAR10. The performance of MNTD is fairly non-robust based on our extensive reproduction evaluations when we test it against conventional backdoored models, in particular, $M_{\rm bd}$ in our case. Nonetheless, we apply it to test $M_{\rm rm}$ and $\widetilde{M}_{\rm rm}$. Note that MNTD uses AUC as a main metric, which assumes that a number of backdoored models and benign models trained over the same task/dataset are under test. This is somehow unrealistic because the model-under-test given the same task in practice is usually the only one model (i.e., benign or backdoored). 

\begin{table*}
\centering 
\caption{Thresholds of each meta-classifier and its evaluation results for PQ backdoor attacks. Five meta-classifiers in total.}
\resizebox{0.90 \textwidth}{!}
{
\begin{tabular}{c | c | c | c | c || c | c | c} 
\toprule 
\multirow{3} * {\begin{tabular}{@{}c@{}} meta-classifier \\ No. \end{tabular}}
& \multirow{3} * {Threshold}
& \multicolumn{3}{c ||}{$M_{\rm rm}$}
& \multicolumn{3}{c}{$\widetilde{M}_{\rm rm}$} \\
\cline{3-8}
&& \begin{tabular}{@{}c@{}} Median of \\ test model scores \end{tabular}
& Average AUC & Accuracy 
& \begin{tabular}{@{}c@{}} Median of \\ test model scores \end{tabular}
& Average AUC & Accuracy \\

\midrule
$1_{\rm th}$ & -2.74 & 4.87 &   \multirow{5} * {49.60\%} &  \multirow{5} * {50.00\%} & 4.75 & \multirow{5} * {46.50\%} & \multirow{5} * {50.00\%}\\
$2_{\rm th}$ & -0.57 & 2.51 &&& 2.54 &\\
$3_{\rm th}$ & 0.13 & 13.27 &&&  13.37 &\\
$4_{\rm th}$ & -2.62 & 6.77 &&&  6.56 &\\
$5_{\rm th}$ & -4.29 & 8.51 &&&  8.44 &\\

\bottomrule
\end{tabular}
}
\label{tab:threshold}
\end{table*}

As the original work does not specify how to evaluate a single model, we instead use the following steps to do so.  
Given a single target model, the meta-classifier outputs a score for that model to measure its maliciousness. However, this score is not constrained by the \textit{sigmoid} activation function, which indicates that the score of the target model does not represent the probability that the model has a backdoor. Thus, \textit{a meta-classifier} uses a threshold to aid in decision making, and it selects the median score of all shadow models in this meta-classifier training set as the threshold. A target model (i.e., model under test) is considered to be backdoored when its score is greater than the threshold; otherwise it is benign. Notably, five meta-classifiers in total with different initialization settings are trained by MNTD to jointly decide whether a backdoor exists in the target model. However, MNTD does not explicitly clarify how the decision is jointly made.  Here, we adopt the common technique of majority voting, that is, if the majority of five meta-classifiers believes the existence of a backdoor, then the model under test is considered to be backdoored.

We train 5 meta-classifiers (using shadow benign models $M_{\rm cl}$ and shadow backdoored models $M_{\rm bd}$) on the CIFAR10,  exactly the same to the source code, to detect a test model set that contains 20 positive samples (benign model) and 20 negative samples (backdoored model). The model structure used for the test set is VGG16. A PQ backdoor attack is used to inject a backdoor into the models as a negative sample (i.e., $M_{\rm rm}$ and $\widetilde{M}_{\rm rm}$, respectively), where all test models are trained until convergence. We report the threshold and median test model scores on each meta-classifier. Following this method described by MNTD, we found that the test model scores were generally higher than the threshold of the meta-classifier, which means that the meta-classifier would consider all the test models to have backdoors even if it is benign. When we adopt the majority voting to evaluate the accuracy and average AUC of the meta-classifiers on the test set, the results are summarized in \autoref{tab:threshold}. We can see that MNTD appears to do random guess. More specifically, similar to ABS, it also fails to detect the backdoor in $\widetilde{M}_{\rm rm}$ that it is ought to detect.

\section{PQ Backdoor Incorporated with Source-specific Backdoor Variants}\label{app:SpecificBackdoor}

In all previous experiments, we intentionally use a simple small trigger and an input-agnostic backdoor attack. This setting, under the threat model of defenses such as Neural Cleanse and STRIP, is tailored for defenses that can readily capture the backdoor behavior if the full-precision model does exhibit. 

It is not non-trivial to perform a stealthier PQ backdoor by incorporating a backdoor variant technique to evade defenses even when they are used to inspect quantized models. When training the backdoored model, 50\% poison rate is used, which includes the cover samples---cover samples are those stamped with triggers but retains their ground-truth labels unchanged~\cite{gao2019strip}. Specifically, we consider a source-specific PQ backdoor and confirm that it can bypass the evaluated defenses, i.e., Neural Cleanse, STRIP, ABS and MNTD. In this setting, the model is ResNet18 and the dataset is CIFAR10. The trigger is $6\times 6$ square located at the bottom right corner. The source class is class 1 and the target class is 0, and all other classes are non-source classes. This means the backdoor effect is only triggered when the trigger is stamped with a sample from the source class~1. The non-source class sample with a trigger does not have the backdoor effect. The attack performance of the source-specific PQ backdoor is detailed in ~\autoref{tab:ssNeuralCleanse} (second row). It is clear that the PQ backdoor is still  effective. 

\begin{figure}[t]
	\centering
	\includegraphics[trim=0 0 0 0,clip,width=0.45\textwidth]{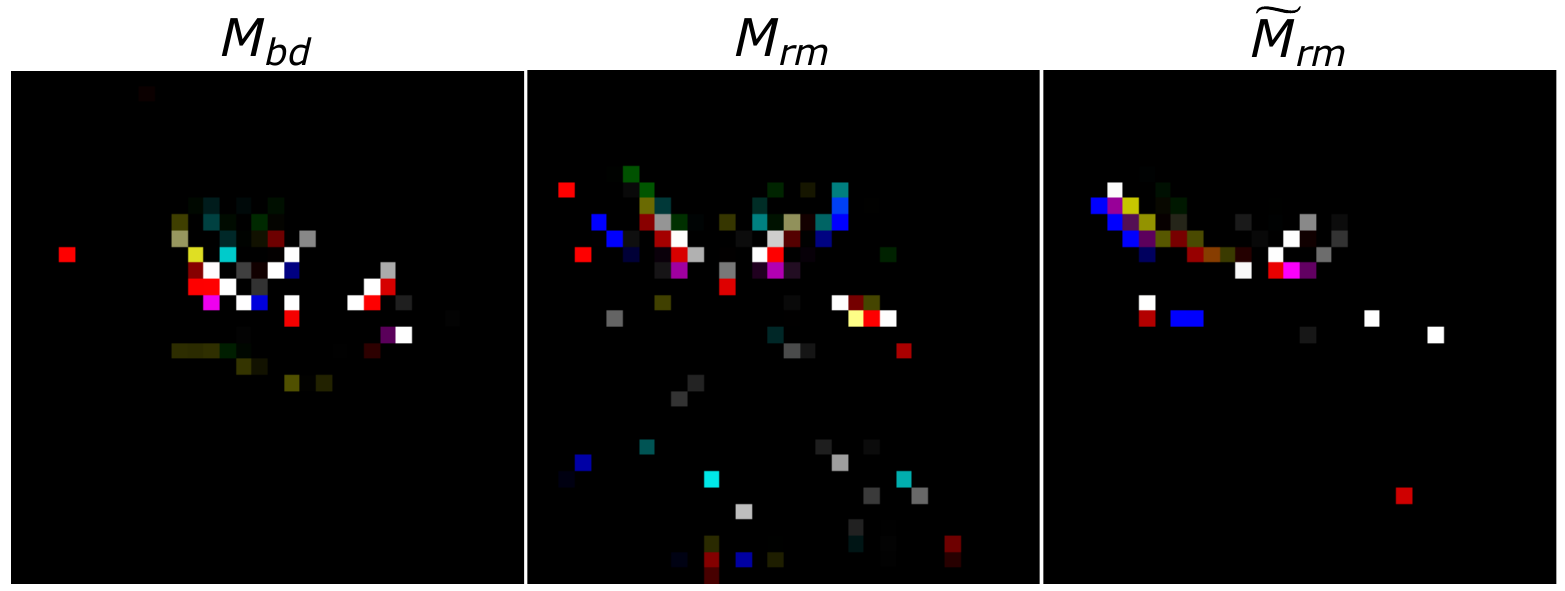}
	\caption{Reverse-engineered triggers of Neural Cleanse for source-specific PQ backdoor attack.}
	\label{fig:ssReverseTrigger}
\end{figure}

\begin{figure}[t]
	\centering
	\includegraphics[trim=0 0 0 0,clip,width=0.45\textwidth]{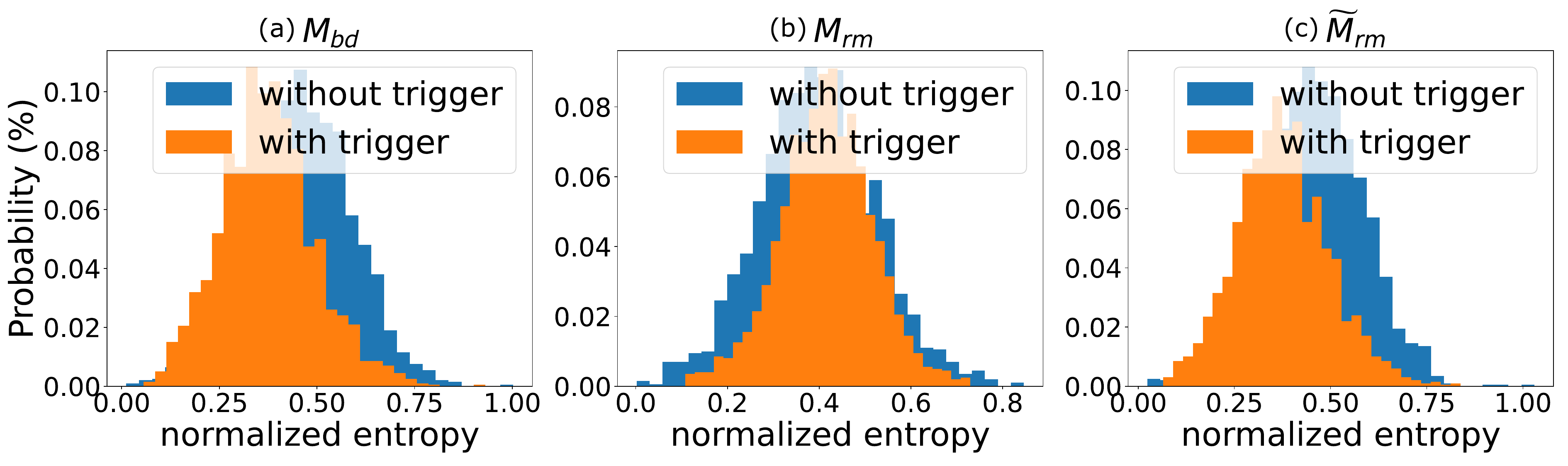}
	\caption{Entropy distribution of the STRIP defense. The source-specific PQ backdoor attack is evaluated.}
	\label{fig:ssSTRIP}
	\vspace{3mm}
\end{figure}

The results of four defenses are summarized as follows.
\begin{itemize}
    \item {\bf Neural Cleanse.} As detailed in ~\autoref{tab:ssNeuralCleanse} (third row), we can see the anomaly index of backdoored model regardless of full-precision $M_{\rm rm}$ or quantized $\widetilde{M}_{\rm rm}$ is lower than 2.0, which means the backdoor is undetected. ~\autoref{fig:ssReverseTrigger} shows reverse-engineered triggers, which are erroneous throughout  backdoored models.
    \item {\bf ABS.} As detailed in ~\autoref{tab:ssNeuralCleanse} (fourth row), the Max REASR of backdoored models are close to clean models (both lower than the threshold of 0.88), which cannot be distinguished by ABS.
    \item {\bf MNTD.} As detailed in ~\autoref{tab:ssNeuralCleanse} (last row), the AUC is close to 50\% in all cases, which implies that MNTD determines the backdoored model by random guess. 
    \item {\bf STRIP.} As shown in ~\autoref{fig:ssSTRIP}, the entropy distributions of the trigger inputs and normal inputs given the backoored models heavily overlap, which indicates that STRIP has failed.
\end{itemize}

In summary, the PQ backdoor using advanced backdoor variants trivially bypasses the 
model inspection against both the front-end full precision model and the quantized model.

\end{document}